\documentclass[10pt, a4paper]{article}

\title{\LARGE Predicting Day-Ahead Stock Returns using Search Engine Query Volumes\\
	   \Large An Application of Gradient Boosted Decision Trees to the S\&P 100}
   
\date{May, 2022}

\author{
	Christopher Bockel-Rickermann, KU Leuven
	\thanks{Research Centre for Information Systems Engineering (LIRIS)}
}

	\usepackage{fontenc}
	
	\usepackage[english]{babel}
	
	\usepackage{csquotes}
	\usepackage[style=apa, citestyle=apa, backend=biber]{biblatex} 
	\usepackage{amsmath}
	\usepackage{amssymb}
	\usepackage{amsthm}
	\usepackage{amscd}
	\usepackage{amsfonts}
	\usepackage{bm} 
	
	\usepackage{graphicx}   
	\usepackage{booktabs}   
	\usepackage{a4}         
	\usepackage{fancyhdr}   
	\usepackage{longtable}  
	\usepackage{subfigure}
	\usepackage{listings}	
	\usepackage[usenames,dvipsnames]{xcolor}	
	\definecolor{KULblue}{RGB}{29,141,176}
	\usepackage{enumerate}	
	\usepackage{layout}		
	\usepackage{tabularx}	
	\usepackage{array}		
	\usepackage{wasysym}	
	\usepackage{fancybox}	
	\usepackage{rotating}	
	\usepackage{slashbox} 	
	\usepackage{multirow}	
	\usepackage{adjustbox}	
	\usepackage{ragged2e}	
	\usepackage[font=small,labelfont=bf,justification=justified,singlelinecheck=no,parskip=5pt,margin=0.2cm]{caption}	
	\usepackage{arydshln}	
	\usepackage{setspace}	
	\usepackage[left=2.54cm, right=2.54cm, right=2.54cm, top=2.54cm]{geometry}
	\usepackage{titlesec}	
	\titleformat{\subsubsection}{}{\thesubsubsection}{1em}{\itshape}
	\usepackage{lipsum}
	\usepackage{float}		
	
	\makeatletter
	\def\nobreakhline{%
		\noalign{\ifnum0=`}\fi
		\penalty\@M
		\futurelet\@let@token\LT@@nobreakhline}
	\def\LT@@nobreakhline{
		\ifx\@let@token\hline
		\global\let\@gtempa\@gobble
		\gdef\LT@sep{\penalty\@M\vskip\doublerulesep}
		\else
		\global\let\@gtempa\@empty
		\gdef\LT@sep{\penalty\@M\vskip-\arrayrulewidth}
		\fi
		\ifnum0=`{\fi}%
		\multispan\LT@cols
		\unskip\leaders\hrule\@height\arrayrulewidth\hfill\cr
		\noalign{\LT@sep}%
		\multispan\LT@cols
		\unskip\leaders\hrule\@height\arrayrulewidth\hfill\cr
		\noalign{\penalty\@M}%
		\@gtempa}
	\makeatother
	
	\usepackage[absolute,overlay]{textpos}
\begin{document}
		\maketitle
		\begin{abstract}
			The internet has changed the way we live, work and take decisions. As it is the major modern resource for research, detailed data on internet usage exhibits vast amounts of behavioral information. This paper aims to answer the question whether this information can be facilitated to predict future returns of stocks on financial capital markets. In an empirical analysis it implements gradient boosted decision trees to learn relationships between abnormal returns of stocks within the S\&P 100 index and lagged predictors derived from historical financial data, as well as search term query volumes on the internet search engine Google. Models predict the occurrence of day-ahead stock returns in excess of the index median. On a time frame from 2005 to 2017, all disparate datasets exhibit valuable information. Evaluated models have average areas under the receiver operating characteristic between 54.2\% and 56.7\%, clearly indicating a classification better than random guessing. Implementing a simple statistical arbitrage strategy, models are used to create daily trading portfolios of ten stocks and result in annual performances of more than 57\% before transaction costs. With ensembles of different data sets topping up the performance ranking, the results further question the weak form and semi-strong form efficiency of modern financial capital markets. Even though transaction costs are not included, the approach adds to the existing literature. It gives guidance on how to use and transform data on internet usage behavior for financial and economic modeling and forecasting.\\[0.5cm]
		\end{abstract}
	
	\section{Introduction\label{Sec:Intro}}
	
	
	A study by the German \textcite{FederalAgencyforCivicEducation.2017} shows that global stock trading volumes increased more than twelve-fold in just ten years between 2005 and 2015. Trading on financial capital markets gained increasing public awareness and importance for hedging risks and financial investment. Professional traders and money managers continuously try to figure out market anomalies, inefficiencies and promising stocks to create returns higher than the average sentiment and to subsequently ``beat the market". Technical stock traders thereby rely on the assumption of weak and semi-strong form inefficiency of markets which means that publicly available information, such as historical stock prices or trading volumes, contain valuable information about the future behavior of the market \parencite{Malkiel.1970}. 
	
	
	\textcite{Krauss.2017} examined the weak form efficiency of financial capital markets by applying modern statistical learning techniques to an empirical set of financial data and successfully proved the historical existence of arbitrage opportunities in the S\&P 500 index. Over time, diminishing returns have proven that market participants started to exploit these investment opportunities with the emergence of widely accessible computation power and research in the field of data analytics. Still, \textcite{Grossman.1980} postulate that markets cannot be informationally efficient when information processing and aggregation is costly. This paper aims to find further evidence for that hypothesis by examining disparate historical data sources.
	
	
	Most importantly, it will focus on the use of information related to user behavior on the internet, namely search engine query volumes. \textcite{Preis.2012} claim that internet users seek information about actions and events in the future rather than the past. If this is true, the specific search behavior of internet users might be a proxy for their future actions. Therefore, the following study tries to answer the question, whether appropriately transformed data on internet user behavior can be used to predict future performances of stocks, subsequently judging the semi-strong form efficiency of modern financial capital markets.
	
	
	For this purpose, a statistical arbitrage strategy is implemented, roughly following the insights from \textcite{Krauss.2017}. Investing in the S\&P 100 index, gradient boosted decision trees are trained to predict directional day-ahead stock returns in the period between 2005 and 2017. The statistical learners use three independent sources of input data: Historical stock returns, relative search volumes for specified terms on the internet search engine Google and additional generic information. Insights from \textcite{Preis.2012, Preis.2013}, as well as \textcite{Choi.2012} are used to transform input data into explanatory variables for model training and prediction making. Each day the five stocks with the highest, as well as the lowest probability of outperforming the index median return are traded. After an empirical simulation of the approach, strategies are evaluated by their generated returns and models are examined to evaluate whether they have a predictive capacity or not. The best performing model, using all of the input data, creates a cumulative performance of more than 1500\% over 11 years of trading. Moreover, all input data appears to have non-zero influence in the models. As a single origin of explanatory variables, alternative data sources are not enough to create returns in excess of the general index. However, in combining separate data sources to ensembles, model quality is substantially increased and cumulative returns are multiples of the performance of the S\&P 100.
	
	
	By exploiting information on internet usage for the prediction of financial capital markets, this paper aims to make several contributions. First, using search engine data for the prediction of stock returns is a novel approach in the existing academic literature. Preceding studies primarily used this kind of information for the prediction of trading volumes. This additional application of search engine data supports the claims of \textcite{Ettredge.2005} and \textcite{Choi.2012} about its relevance for academics and business for short-term economic forecasting. Second, the research adds to the current literature on statistical arbitrage strategies, most dominantly \textcite{Krauss.2017}. In applying their approach to a different index, results are tested for possible generalizations. What is more, this paper presents a meaningful extension to their approach, extracting information different from classic financial data. Third, the results enable to judge current levels of efficiency on financial capital markets, by classifying the data accordingly to the widely accepted postulates of \textcite{Malkiel.1970}.
	
	The paper, therefore, adds value to both the financial industry and related academic research. Testing the value of data on internet usage behavior, the results add new ways for investors to trade on the market. Additionally, the approach proofs to be meaningful and might help researchers and practitioners to facilitate, transform and understand information on search engine usage. It hereby adds to settings different from investment decision making. Related fields are, among others, economic forecasting and modeling, behavioral finance and psychology.
	
	
	The remaining sections of the paper are organized as follows: In Section~\ref{Sec:LitRev} relevant literature is discussed and reviewed. Section~\ref{Sec:Data and Proc} presents the data used for the research, as well as an overview of the software used for the analysis. Section~\ref{Sec:Method} summarizes the general approach of creating explanatory and dependent variables, as well as the procedure of model training, implementation of trading and model evaluation. Section~\ref{Sec:Results} presents the results of the empirical research and performance of statistical models, with an ensuing discussion in Section~\ref{Sec:Discus}. Section~\ref{Sec:Concl} concludes with a final closing summary and possible extensions for future research.
	
	\section{Literature Review\label{Sec:LitRev}}
	
	The literature review covers two topics that are most relevant to this paper. First, it provides an overview of the literature on statistical arbitrage specifying the investment strategy implemented in Section 4. Second, it summarizes the main findings about the use of search engine data in economic forecasting and finance.
	\subsection{Statistical Arbitrage}
	
	The term ``Statistical Arbitrage" on financial capital markets emerged in the mid-1980s from the increasing tendency of financial professionals to use elaborated statistical and mathematical models to trade stocks \parencite{Gatev.1998}. Complex rationale and computer algorithms replaced gut feeling and intuition. The literature does not define the term unambiguously, but present works agree on significant essential aspects. According to \textcite{Hogan.2004} and \textcite{Avellaneda.2010}, statistical arbitrage includes a group of trading strategies that exploit systematic trading signals and rules, having zero initial costs. More precisely, statistical arbitrage is a positive return on financial markets, which is (a) generated by a procedure where historical information is used to predict the future accurately and (b) not requiring any initial investment. The latter is usually achieved by buying assets that are believed to gain in value and short selling assets, that are believed to fall in value.
	
	The existing research mostly varies in the approach of information acquisition and processing, as well as in the covered time horizon. One of the earliest papers on statistical arbitrage is the work of \textcite{Gatev.1998} on the implementation of a strategy called ``pairs trading". Pairs trading relies on the empirical observation that specific pairs of stocks show a similar price path. It assumes that current deviations from this path will reverse in the future. \textcite{Gatev.1998} detected pairs by minimizing the sum of squared deviations of historical stock returns on the US-American stock market. When they found a pair, they bought the stock that negatively deviated from the common path and sold short the stock that deviated positively. While overall returns were significantly larger than zero, sub-period returns were declining over time. Due to the simplicity of the approach these results are remarkable.
	
	\textcite{Alexander.2005} varied the approach of pairs trading by identifying stocks within a fixed investment universe that show a high level of empiric co-integration, rather than similar historical returns as in \textcite{Gatev.1998}. Stated returns appear both to be higher and have less variance than based on the approach of \textcite{Gatev.1998}, suggesting that their approach is more stable. The cointegration approach to statistical arbitrage was also used in more recent studies by \textcite{Avellaneda.2010} and \textcite{Caldeira.2013}.
	
	As opposed to the idea of identifying single pairs of stocks, an increasing amount of research focusses on the identification of more extensive trading portfolios, usually by modeling bigger sets of data for understanding and predicting higher numbers of stock returns. 
	
	\textcite{Hogan.2004}, for example, built portfolios of stocks based on historical relationships between stock returns and fundamental figures, such as book-to-market ratios or market capitalisation. In this way, \textcite{Hogan.2004} classified stocks in well performing and poorly performing ones, instead of  forecasting absolute returns. In their choice of fundamental figures they built on the study of \textcite{Fama.1992} on market anomalies based on book-to-market ratios and sizes of companies. Portfolios are built based on stocks with the highest and lowest values for the relevant figures.
	
	More recent studies rely on the emergence and widespread availability of data processing algorithms and computation power, exploiting academic improvements in data analytics and data science:
	
	\textcite{Huck.2009} applied deep neural networks to forecast week-ahead returns of constituents of the S\&P 100 index. Trading portfolios are built based on a certain number of the most and least promising stocks on the index, focusing on the time interval between 1992 and 2006. However, the approach used in \textcite{Huck.2009} is questionable as it did not account for variations in the index composition over time. In contrast, the study only works with the set of constant constituents of the index, which raises questions about possible survivorship bias in the results. 
	
	Most important for the proceeding of this paper is the study of \textcite{Krauss.2017}, who built upon the insights of \textcite{Huck.2009}. \textcite{Krauss.2017} predicted stock-wise directional outperformance on the S\&P 500. Motivated by \textcite{Takeuchi.2013} and \textcite{Dixon.2015}, they used historical stock price movements to predict, whether or not a stock is going to outperform the median return of the index on the following day on a time frame from 1992 to 2015. \textcite{Krauss.2017} evaluated the performance of gradient boosted decision trees, random forest classifiers and artificial neural networks for this purpose. Results show significant returns over the presented period, suggesting the superiority of the tree based models over the artificial neural network, an insight that is surprising in light of the increasing popularity of neural networks both in academia and business.
	
	\subsection{Facilitating Internet Usage Data}
	
	As vital interfaces to most parts of the publicly accessible internet, search engines have the potential to collect vast amounts of behavioral information about their users and associated topics. Facilitating this knowledge is an emerging topic in academia and economics.
	
	One of the earliest studies on the potential importance of internet usage data, \textcite{Ettredge.2005} provided early evidence for existing relationships between user behavior on major global search engines and real-world indicators. Specifically, they discovered the positive correlation of search query volumes and unemployment rates.
	
	Subsequently, \textcite{Choi.2012} showed that data from the Google Trends database on search query volumes could be used for forecasting events in the nearer future. In an empirical study using historical data on search volumes, they accurately predicted numbers of automotive sales in the USA. Furthermore, they postulated that especially actions that are planned in advance by action takers or that require a degree of planning can be forecast using search query volumes.  An interesting claim that is supported by \textcite{Preis.2012}, who showed with an empirical set of search queries that internet users search for events in the future, rather than for events in the past. This insight links the topic of search engine query volumes to financial markets and investing, where actors on the market usually spend large amounts of time and workforce in researching potential investments and observing the latest stock market-related news.
	
	\textcite{Mao.2011} evaluated relationships between distinct sources of internet related data and various financial statistics. Predominantly, they used numbers of tweets on a particular topic on the social media platform Twitter as well as Google search volumes to predict trading volumes and the level of the NASDAQ index. Results are supporting the applicability of internet related data to finance. Using search volumes from the search engine Yahoo!, \textcite{Bordino.2012} broadened the insights of \textcite{Mao.2011} and showed that search volumes can indeed predict trading volumes on a lower hierarchical level, predicting stock-wise daily trading volumes for the 100 constituents of the NASDAQ-100 index. Applying search volume data to individual assets on financial capital markets also marks a leap in the possible granularity of the approach, which is insightful for many other fields than finance.
	
	Meanwhile, there is little literature about the actual facilitation of search volume data for investing and trading on financial markets. \textcite{Preis.2013} investigated simple strategies, in which they invested in the general stock market, when search volumes for a defined search term increase and divest when volumes decrease. Their study showed that this simple approach can indeed create substantial amounts of return over the period considered. They, however, do not answer the question of how to identify promising search terms, raising questions about possible data snooping and coincidence in their study. \textcite{Weng.2017} focused on forecasting day ahead stock returns in the context of providing an expert system to investors that exploits big data and internet search volumes. However, their results appear to be questionable, as their research focuses only on a minimal amount of data and one stock to predict, namely the US-American technology company Apple. The particular context and limitations of the data, therefore, question possibilities of further generalization of their results.
	
	\section{Data and Processing\label{Sec:Data and Proc}}
	
	\subsection{Data Aggregation\label{Sec:Data and Proc_Data}}
	
	The target index of the empirical study is the S\&P 100. Since its launch in 1983, the index lists 100 of the largest companies by market capitalization traded on US-American exchanges \parencite{SandP.2019}. The choice of the S\&P 100 is primarily motivated by its size, analyst coverage and name recognition of its constituents. \textcite{Huck.2009} claims that the constituents are among the most highly monitored and highly liquid stocks worldwide. Additionally, it is a true sub-sample of the S\&P 500 index, which is the most prominent reference index to test financial modeling, emphasizing the validity of choice. 
	
	The time frame chosen for the analysis are the years from 2005 to 2017, resulting in a total of 13 years covered. This period reflects the younger past only, in order to ensure availability of data, but still covers essential historical events, such as the financial crisis of 2008 and 2009.
	
\begin{figure}[htbp] 
		\centering
		\footnotesize
		\includegraphics[width=0.9\textwidth]{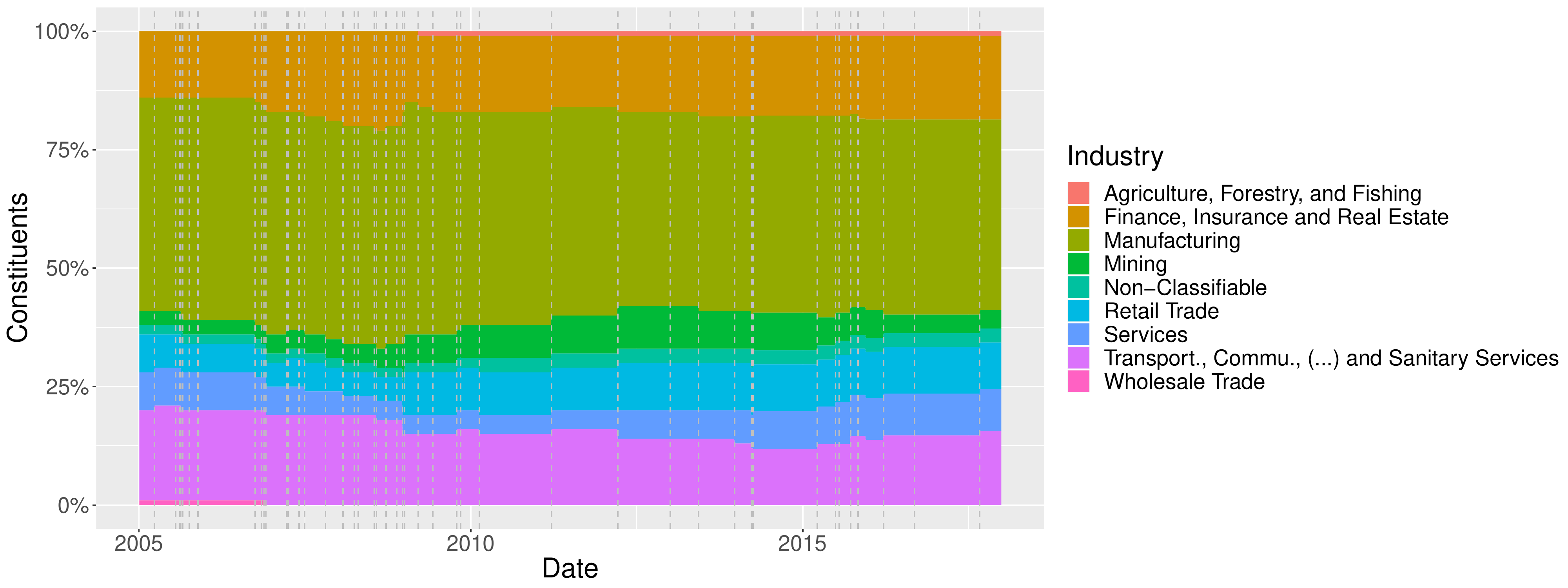}
		\captionsetup{justification=justified} 
		\caption[S\&P 100 Index Composition]{S\&P 100 Index Composition \newline Vertical dashed lines represent a change in the index composition.} 
		\label{fig:Composition}
	\end{figure}

	The composition of the index is changing over time. As stocks are traded continuously, market capitalization change following supply and demand. Thus individual stocks are entering and leaving the index. To account for changes, the empirical index composition is evaluated daily. A list of all historical constituents of the S\&P 100 is downloaded from the Compustat - Capital IQ database, managed by Standard \& Poor's \parencite{Standard.2019}\footnote{The database has been accessed via \textcite{WRDS.2019b}.}. The list provides both the date a company has entered the index and, if applicable, the day on which it has left it, as well as information on the industry, in which a company is operating. Industry information is provided by the corresponding SIC Division, an industry identifier assigned by \textcite{USDL.2019}. From there on, the information is transformed into lists for each trading day, containing all companies that have been a constituent of the index over that entire day. Overall, the analysis considers 170 individual companies that have been part of the index from 2005 to 2017. Figure~\ref{fig:Composition} visualizes the change in composition, depicting both immediate changes in constituents and the relative composition by industries over time. The figure shows that especially in the period before 2010 changes in the index composition occur more frequently than in the period after 2010. The distribution of industries remains relatively stable.
	
	In order to create input data for model training and prediction making (cf. Section~\ref{Sec:Method}), corresponding data for each stock from the constituent lists is downloaded. The data is divided into three major sources of information; historical returns, relative internet search volumes and generic information and considers trading days only. In order to build efficient ensembles of information, the sources are chosen to be uncorrelated \parencite{Ettredge.2005}. Further description follows below:
	
	\subsubsection{Historical Returns\label{Sec:Data and Proc_Data_Hist Ret}}
	
	For each of the historical constituents, their daily holding period returns are downloaded. The time series originate from the CRSP database \parencite{CRSP.2019}\footnote{The database has been accessed via \textcite{WRDS.2019a}.}. The holding period return contains information about all relative capital gains and losses an investor is facing from holding shares of a stock from one day to another. It thereby accounts for any events that are return relevant, such as dividend payments and price changes. Subsequently, the holding period return will be referred to by
	
	\begin{equation}
		R_{t}^{i} = \frac{(P_{t}^{i} - P_{t-1}^{i}) + D_{t}^{i}}{P_{t-1}^{i}},
	\end{equation}
	where $t$ specifies a specific day, on which the stock of company $i$ was traded. $P_{t}^{i}$ is the price of a share of company $i$ at the end of day $t$ and $D_{t}^{i}$ is the potential dividend payment for the stock at day $t$. The choice of historical stock returns for predicting future returns is motivated by the works of \textcite{Krauss.2017} and \textcite{Takeuchi.2013}.
	
	\subsubsection{Internet Search Volumes\label{Sec:Data and Proc_Data_Internet}}
	
	With the extensive availability of internet access in the developed world, the internet provides a continually increasing set of publicly available information to its users. Due to the number of websites with varying importance and quality, research on the internet usually starts on a search engine, enabling to find relevant websites and resources. Among others, \textcite{Choi.2012} have pointed out that the behavior of users on a search engine can be a useful resource for forecasting tasks. If correctly interpreted, search engine data might, therefore, have predictive power for statistical modeling. In the context of financial markets, for example, individuals might interact with the internet first, in order to research a company, before physically engaging with markets by buying or selling shares of a stock. The insights from \textcite{Bordino.2012} and \textcite{Weng.2017} further underline this hypothesis.
	
	Using the above insights, data is collected from the Google Trends database \parencite{GoogleTrends.2019}. According to a study by \textcite{StatCounter.2019}, by the end of 2017  Google obtained a market share of 88.6\% among other search engines in the USA. This high share of the market is making it a valid proxy for the overall behavior on search engines. On its Trends database, Google has published information on relative search volumes for specified search terms since the beginning of 2004. For an individually defined periodicity and time frame, Google calculates the proportion of searches for a certain term compared to the total amount of searches on the engine and re-scales the individual values to a scale from 0 to 100, based on a sampled set of search queries. These time series are subsequently referred to as Search Volume Indexes (SVI). For each of the constituents of the S\&P 100 index, the daily SVI of each of the companies' ticker symbol is downloaded. The ticker symbol is a unique abbreviation of the company name that is used to ease communication about a stock on financial capital markets. Using the ticker symbol aims to measure the behavior of potential traders of a stock more closely, instead of the behavior of agents that are not engaged in financial capital markets. To further decrease bias in the SVI by searches that have not been company related, the functionality of the Trends database is used to limit the SVIs to company-related searches. In this way, ambiguity can be significantly reduced, in case a ticker symbol equals another commonly used word or phrase. All data is collected for search queries in the USA only, motivated by the outcomes of \textcite{Preis.2013}, stating that search data from the USA has a higher predictive value for the performance of US-American stocks than global data. In case of overall low popularity of a search term and insufficient data to calculate the SVI, Google provides the values $0$ or ``$<1$". For consistency, ``$<1$" is likewise transformed to ``$0$".
	
	\begin{figure}[htbp] 
		\centering
		\includegraphics[width=0.6\textwidth]{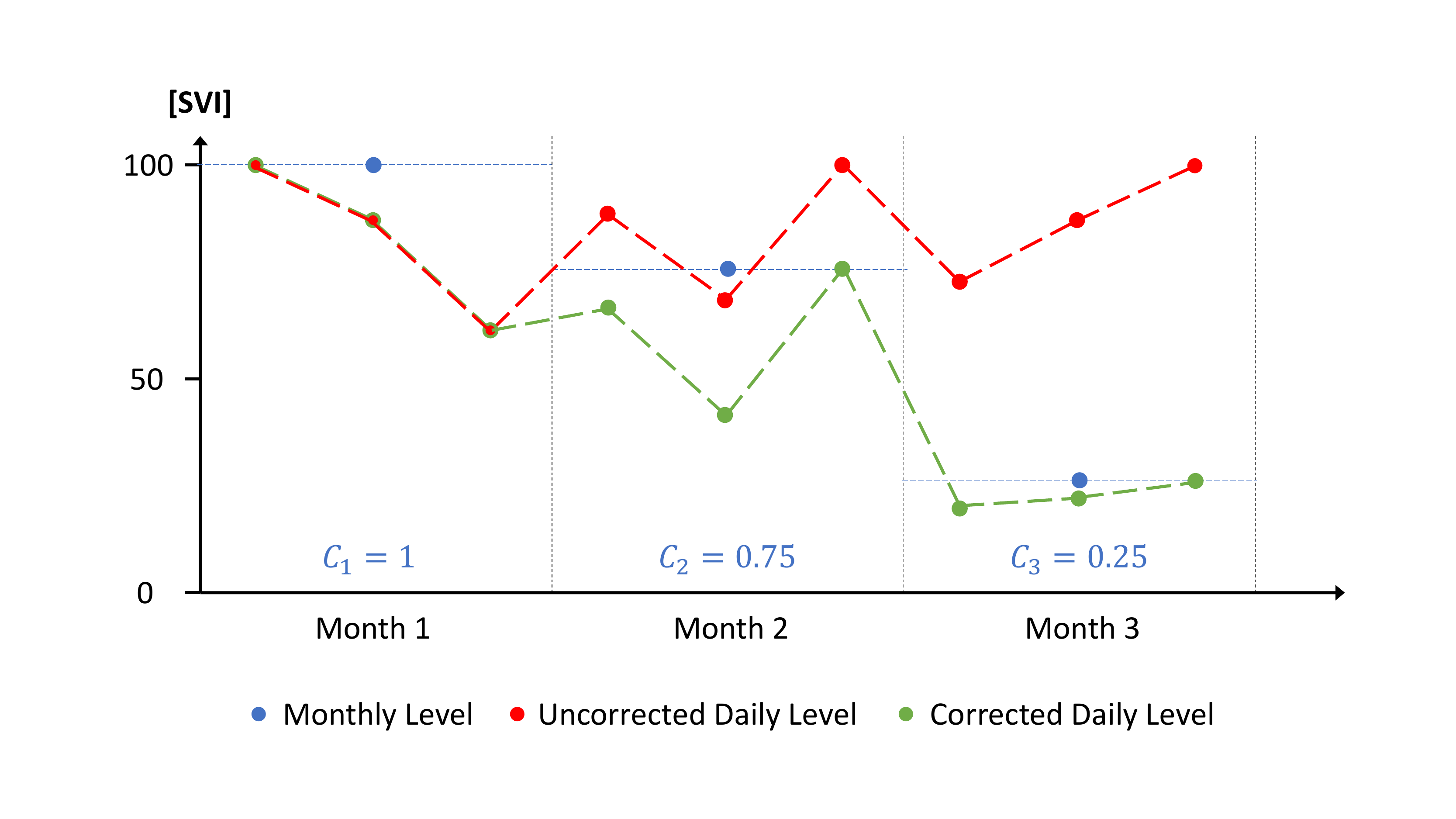}
		\captionsetup{justification=justified} 
		\caption[Visualization of Daily SVI Adjustment]{Visualization of Daily SVI Adjustment} 
	\label{fig:SVI_adjustment}
	\end{figure} 
	
	However, obtaining a daily SVI time series over several years cannot be done in a straight forward fashion. Google only allows for downloading daily SVIs over a period not longer than 90 days. To build a continuous daily time series for more than 90 days, individual smaller series have to be concatenated. As Google re-scales every time series that is downloaded to a 0 to 100 scale, each fragment has to be corrected for its overall monthly level on the entire time frame. This correction is done with the help of the corresponding monthly SVI, which is available for all time frames since the beginning of the database. Figure~\ref{fig:SVI_adjustment} illustrates the approach with fictional data over three months. First, daily time series are downloaded for each month individually. In the figure, daily time series contain three observations each, initially scaled from 0 to 100. These values are depicted in red. Second, a monthly time series for the entire time frame specifies the change in level from month to month. In the figure, the monthly time series is depicted in blue. It can be seen that the monthly time series is downward-trending and that the daily time series is not. To account for the monthly trend, each daily time series is corrected by the relative monthly level. To do so, values are multiplied by $(Monthly Level/100)$. In Month 1, the overall level of the SVI is 100, so the daily time series gets multiplied by 1, analogously months 2 and 3 get multiplied by 0.75 and 0.25, resulting in a corrected daily time series. The corrected time series is matching the overall monthly trend and is depicted in green. Figure~\ref{fig:SVI_adjustment_bonds} in the appendix shows the outcome of the correction for the search term ``bonds".
	
	\begin{figure}[htbp] 
	\centering
	\footnotesize
	\includegraphics[width=0.5\textwidth]{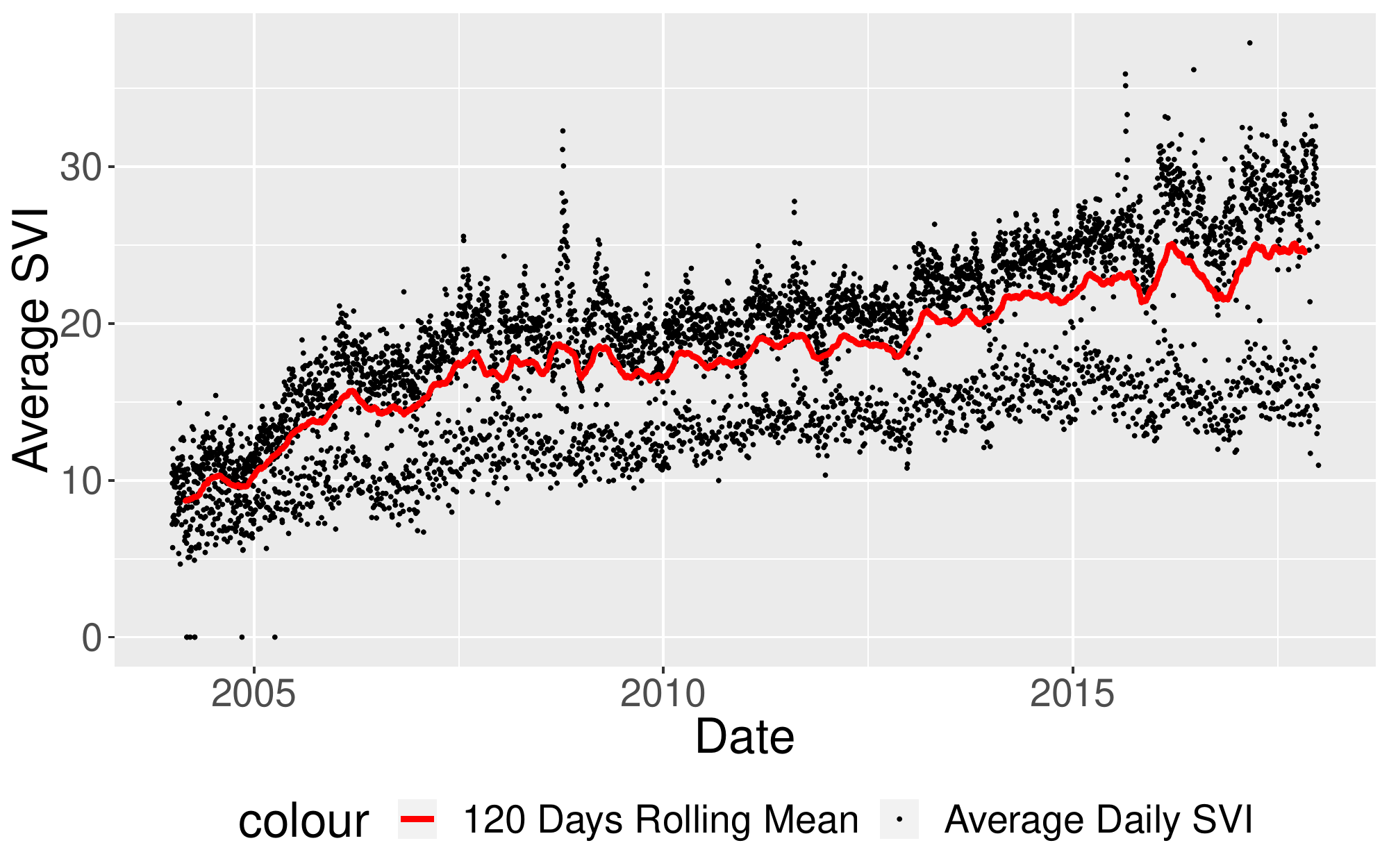}
	\captionsetup{justification=justified} 
	\caption[Mean SVI of S\&P 100 constituent ticker symbols]{Mean SVI of S\&P 100 constituent ticker symbols} 
	\label{fig:Mean_SVI}
	\end{figure}
	
	Examining the resulting data on internet search volumes provides essential insights. The data is very noisy. Figure~\ref{fig:Mean_SVI} plots the mean SVI over all index constituents on each day from 2005 to 2017. Even averaging over all constituents is not reducing noise and variance of the time series. Plotting the 120-day moving average reduces the high variance in the data and shows a positive trend in the average SVI, from a relatively low value of 10 in 2005, to about 25 in 2017.
	
\begin{figure}[htbp] 
	\centering
	\footnotesize
	\includegraphics[width=0.5\textwidth]{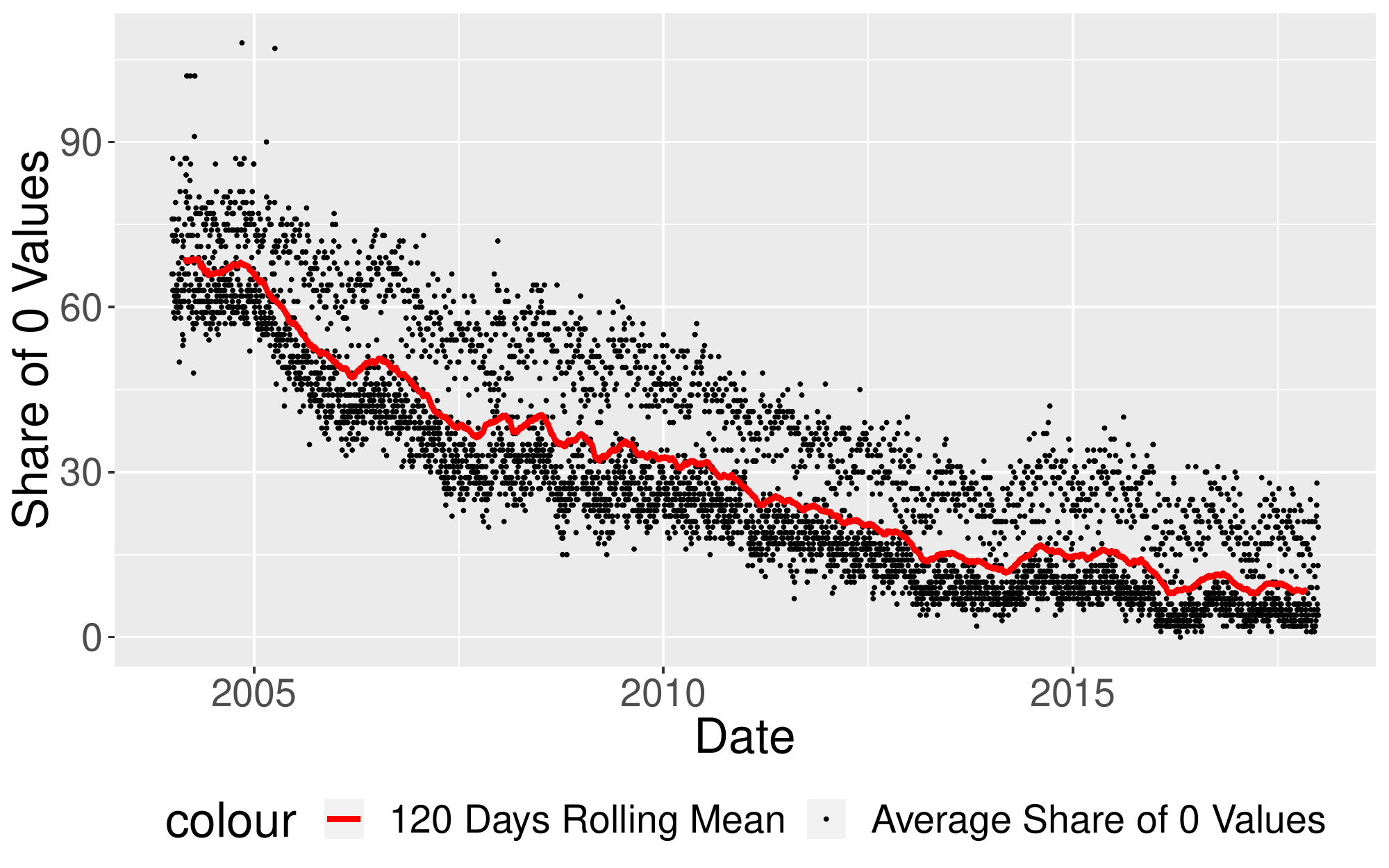}
	\captionsetup{justification=justified} 
	\caption[Share of zero-value SVIs per day]{Share of zero-value SVIs per day} 
	\label{fig:0_values_SVI}
	\end{figure}

	The overall increase on the average SVI happens analogously to a higher amount of available data. Figure~\ref{fig:0_values_SVI} shows the share of 0-values in the SVIs of index constituents on a certain date. Just as the overall average, this value exhibits high variance. The 120-day rolling mean shows that especially in the earlier history, an average of over 60\% of the SVIs is missing reliable information from the database.
	
	\subsubsection{Generic Information\label{Sec:Data and Proc_Data_Generic}}
	
	As a final source of data, three additional sets of information are added to the historical returns and internet search volumes. 
	
	First, the industry information considered in Section~\ref{Sec:Data and Proc_Data} is saved. Using this piece of information for prediction making is motivated by the empirical difference in both the size of returns and return variances across stocks from different backgrounds. Table~\ref{tbl:IndustryOverview} shows statistics on the industry composition of the index, as well as average monthly returns and standard deviations for the period from 2005 to 2017. For example, stocks from the mining industry on average account for the same number of shares in the S\&P 100 as companies from the services industry, but achieve a lower return. Enabling a statistical model to use this information, might result in better accuracy for prediction making, especially, as models are trained on the entire panel of data, instead of for individual stocks separately.  
	
	\begin{table}[!htbp] 
	\caption[Industry Overview S\&P 100]{Industry Overview S\&P 100 \newline The table summarizes monthly mean returns by industry, as proposed by the companies' SIC codes and corresponding SIC divisions for the constituents of the S\&P 100 index between 2005 and 2017.}
	\label{tbl:IndustryOverview}
	\centering
	\footnotesize
	\begin{adjustbox}{max width=15.2cm}
	\renewcommand{\arraystretch}{0.8}
	\begin{tabular}{@{\extracolsep{5pt}}lccc} 
	\\[-1.0ex]\hline 
	\hline \\[-1.0ex] 
	Industry (SIC Division) & Mean No. Stocks & Mean Return [\%] & Mean Standard Deviation [\%]\\
	\hline \\[-1.0ex] 
	Agriculture, Forestry, And Fishing & 0.99 & 1.74 & 8.90\\[0.3ex]
	\hdashline\\[-1.00ex]
	Mining & 5.78 & 1.18 & 10.92\\[0.3ex]
	\hdashline\\[-1.00ex]
	Construction & - & - & -\\[0.3ex]
	\hdashline\\[-1.00ex]
	Manufacturing & 43.76 & 1.08 & 8.33\\[0.3ex]
	\hdashline\\[-1.00ex]
	Transportation, Communications, Electric, Gas, & \multirow{2}{*}{15.78} & \multirow{2}{*}{1.05} & \multirow{2}{*}{8.42}\\[-0.1ex]
	And Sanitary Services &  &  &\\[0.3ex]
	\hdashline\\[-1.00ex]
	Wholesale Trade & 1.00 & 1.24 & 17.84\\[0.3ex]
	\hdashline\\[-1.00ex]
	Retail Trade & 8.43 & 1.12 & 8.80\\[0.3ex]
	\hdashline\\[-1.00ex]
	Finance, Insurance, And Real Estate & 16.50 & 1.15 & 13.18\\[0.3ex]
	\hdashline\\[-1.00ex]
	Services & 6.34 & 1.44 & 9.84\\[0.3ex]
	\hdashline\\[-1.00ex]
	Public Administration & - & - & -\\[0.3ex]
	\hdashline\\[-1.00ex]
	Non-classifiable & 2.59 & 0.85 & 6.98\\
	\hline \\[-1.0ex]
	\end{tabular}
	\end{adjustbox}
	\end{table}       
	
	Second, the returns also appear to vary between different weekdays (cf. Table~\ref{tbl:WeekdayOverview} in the appendix). Weekdays are therefore considered as a further predictor within the modelling (cf. Section~\ref{Sec:Method}).
	
	Third, the daily trading volume of a stock in US\$, calculated as the number of individual stocks traded multiplied by the price of the stock at the end of the corresponding trading day. The trading volume is an important indicator of financial capital markets. It might indicate how much attention a stock is attracting. Prominently used by traders, the trading volume might as well contain valuable information for prediction price movements. It is subsequently referred to as $V_{t}^{i}$, where $t$ specifies a certain day, on which company $i$ was traded. Stating the trading volume in the total amount of US\$ serves the purpose of overcoming the problem of stock splits and other events that are price-relevant, but not return-relevant. In case of a stock split, each share of a stock gets automatically divided into $n$ new shares, each worth $1/n$ the value of the original share. This does not change the market capitalization of a company or the value of an investor's portfolio; however the total number of shares traded usually multiplies by the factor $n$, as traders do trades in terms of absolute monetary value, rather than a specific number of shares.
	\subsection{Data Processing\label{Sec:Data and Proc_Proc}}
	The complete statistical analysis is conducted in the programming language ``R" \parencite{R.2018}, including processing the data, model training and evaluation. The approach is more closely described in Section~\ref{Sec:Method}. Moreover, the approach relies on functionalities from different R packages by other authors. Data subsetting and formatting is done using the ``data.table" package \parencite{data.table.2019}. Statistical learners are implemented using the ``gbm" package for gradient boosted trees \parencite{gbm.2019}. Tuning parameters of statistical models are optimized based on their receiver operating characteristic (ROC), computed using the ``pROC" package \parencite{pROC.2011}. Performance testing is conducted using functionalities from the ``performanceAnalytics" package \parencite{performanceAnalytics.2018}. Visualizations have been created using the ``ggplot2" package \parencite{ggplot2.2016}. Downloads from the Google Trends database (cf. Section~\ref{Sec:Data and Proc_Data_Internet}) are also automated using ``R", as well as the ``gtrendsR" package \parencite{GtrendsR.2018}.
	
	\section{Method\label{Sec:Method}}
	
	The statistical analysis is carried out in four successive steps, following \textcite{Krauss.2017}. In the first step, the time frame of the analysis is split into individual sub-periods, each composed of a training set and a test set. Carrying forward, for each of the sub-periods the necessary features are created to do model training and evaluation. In a third step, statistical models are trained individually on the training sets. Finally, the models are used to make predictions on the test sets and results are used for trading. Each of the steps is further outlined below.
	
	\subsection{Subsetting\label{Sec:Method_Subsetting}}
	
\begin{figure}[htbp] 
		\centering
		\footnotesize
		\includegraphics[width=0.8\textwidth]{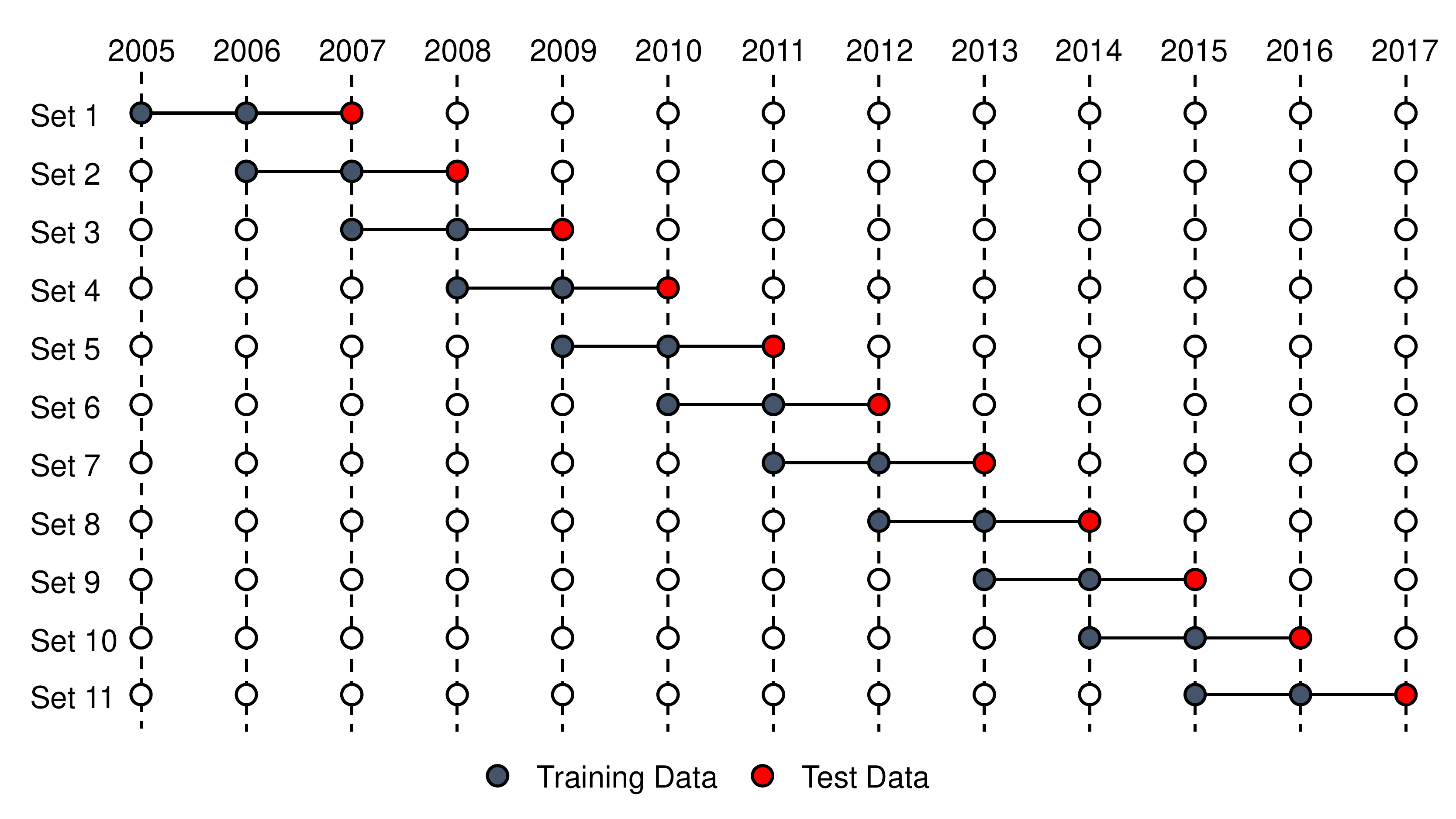}
		\captionsetup{justification=justified} 
		\caption[Subsetting Scheme - Training and Test Sets]{Subsetting Scheme - Training and Test Sets} 
		\label{fig:Subsetting}
	\end{figure}

	The entire time frame of the analysis is divided into subsets of three years of length. The first two years of a subset function as a training set, in order to build statistical models. The third year is used as a test set for prediction making. Only test sets are non-overlapping, in order to mimic real-world behavior of continuously retraining models after defined periods and continuous trading from day to day. This approach results in a total of 11 sub-periods, visualized in Figure~\ref{fig:Subsetting}. Index constituent lists (cf. Section~\ref{Sec:Data and Proc_Data}) are then used to reconstruct the empirical index composition over the sub-periods. On average, training sets consist of 503.5 days and test sets consist of 251.7 days.
	
	\subsection{Feature Creation\label{Sec:Method_Features}}
	
	In order to train models and make predictions on the sub-periods, first, explanatory variables are created, followed by the creation of response variables. Response variables relate to a certain day $t$, therefore, explanatory variables are created such that they only contain information from days prior to $t$. 
	
	\subsubsection{Explanatory Variables\label{Sec:Method_Features_X}}
	
	Explanatory variables are created individually for the different data categories presented in Section~\ref{Sec:Data and Proc_Data}. For the historical returns, historical cumulative returns are calculated for each day $t$ and stock $i$ by using the information on daily holding period returns. The explanatory variables are created as the cumulative return of a stock $i$ over the last $d$ days prior to $t$. The variables are formally defined as
	
	\begin{equation}
	CR_{t,d}^{i}=\prod_{k=1}^{d}(1+R_{t-k}^{i}).
	\end{equation}
	Taking the insights of \textcite{Takeuchi.2013} and \textcite{Krauss.2017}, cumulative returns are created for all $d \in \{1, 2, 3, 4, 5, 10, 15, 20, 40, 60, 120, 180, 240\}$. Thus granularity is continuously reduced when creating features from observations dating further into the past. Observations, where less than 240 days of historical returns are available, are removed from the data set.
	
	The information about internet search volumes is handled by calculating historical changes in the SVI of a specific company $i$. This approach is motivated by \textcite{Preis.2012} and the evidence that individuals seek information on the internet about the future, including exogenous events and individual behavior. As it is unknown, whether and how this holds for behavior on financial capital markets, features are created from the daily percentage change in the SVI on a specific day preceding the observed day $t$. More precisely the variable measures the percentage change of the SVI of company $i$ from day $t-(d+1)$ to the following day $t-d$. The variables are formally defined as
	
	\begin{equation}
	\Delta SVI_{t,d}^{i} = \begin{cases}
		\dfrac{SVI_{t-d}^{i} - SVI_{t-d-1}^{i}}{SVI_{t-d-1}^{i}} & ,SVI_{t-d-1}^{i} > 0\\
		0 & ,SVI_{t-d-1}^{i}=SVI_{t-d}^{i}=0\\
		99 & , else.
	\end{cases}
	\end{equation}
	The conditional handling of 0 values is necessary due to the structure of the data depicted in Section~\ref{Sec:Data and Proc_Data_Internet}. The approach prevents missing or false predictors when training the models. When data is missing for two days in a row $\Delta SVI_{t,d}^{i}$ is set to 0 to indicate that there is no change of query volumes. When at one day the SVI is 0 and increases to $>0$ on the following day $\Delta SVI_{t,d}^{i}$ is set to 99, an arbitrarily high value, to indicate a positive change. This handling enables decision trees to separate these observations from the rest of the population. Features are created for all $d \in \{1, 2, 3, 4, 5\}$, taking into consideration the daily changes in SVI of a company within the last trading week.
	
	
	For the generic information set, a categorical variable is created both for the weekday of an observation, namely $W_{t} \in \{Mon, Tue, Wed, Thu, Fri\}$, as well as for the industry, in which a company operates, referred to as $I^{i}$, with levels accordingly to the industry mapping in Table~\ref{tbl:IndustryOverview}. Likewise to the processing of SVIs, the daily trading volume per company is handled by calculating past relative changes. Therefore, the change in trading volume of company $i$ on the last day prior to $t$ is considered which is defined as
	\begin{equation}
	\Delta V_{t,1}^{i} = \dfrac{V_{t-1}^{i} - V_{t-2}^{i}}{V_{t-2}^{i}}.
	\end{equation}
	
	\subsubsection{Response Variables\label{Sec:Method_Features_Y}}
	
	The response variable notes, whether or not a stock's return beats the index' median return on day $t$. Predicting whether or not a stock is beating an index median is a common approach in the statistical arbitrage literature. Furthermore, defining a classification problem is motivated by the findings of \textcite{Enke.2005} and \textcite{Leung.2000}, claiming that classification problems result in more robust models and predictions. The dependent variable on a specific day $t$ is therefore defined as
	
	\begin{equation}
	Y_{t}^{i}=\begin{cases}
		1 & ,R_{t}^{i} > median(R_{\{K_{t}\}})\\
		0 & ,else
	\end{cases},
	\end{equation}
	for a company $i$, where $R_{\{K_{t}\}}$ is the set of index constituent returns on day $t$ that have remained in the data after the creation of the explanatory variables. Observations, where $R_{t}^{i}$ is not available, are removed from the dataset.

	\subsubsection{Feature Creation Summary\label{Sec:Method_Features_summary}}
	
	The above approach results in training and test sets with one response and three different sets of explanatory variables, summarized in Table~\ref{tbl:PredictorSets}. In total, the dimensions of the data set are (326,589 x 22), covering 3,272 days. In total 2,183 observations have been removed, because they were missing data in the predictor variables. Therefore, about 0.66\% of the observations are missing, a value that is small enough to assume an unbiased analysis.
	
	\begin{table}[!htbp] 
	\caption[Summary of Predictor Sets]{Summary of Predictor Sets}
	\label{tbl:PredictorSets}
	\centering
	\footnotesize
	\begin{adjustbox}{max width=15.2cm}
		\renewcommand{\arraystretch}{0.8}
		\begin{tabular}{@{\extracolsep{5pt}}lcl} 
			\\[-1.0ex]\hline 
			\hline \\[-1.0ex] 
			Predictor Set & Abbreviation & Components \\[+0.5ex]
			\hline \\[-1.0ex] 
			\multirow{2}{*}{Historical Returns} & \multirow{2}{*}{CR} & $\{$ $CR_{t,1}^{i}$, $CR_{t,2}^{i}$, $CR_{t,3}^{i}$, $CR_{t,4}^{i}$, $CR_{t,5}^{i}$, $CR_{t,10}^{i}$, $CR_{t,15}^{i}$\\
			& &  $CR_{t,20}^{i}$, $CR_{t,40}^{i}$, $CR_{t,60}^{i}$, $CR_{t,120}^{i}$, $CR_{t,180}^{i}$, $CR_{t,240}^{i}$ $\}$ \\[0.80ex]
			\hdashline\\[-1.00ex]
			Internet Search Volumes & SVI & $\{$ $\Delta SVI_{t,1}^{i}$, $\Delta SVI_{t,2}^{i}$, $\Delta SVI_{t,3}^{i}$, $\Delta SVI_{t,4}^{i}$, $\Delta SVI_{t,5}^{i}$ $\}$\\[0.80ex]
			\hdashline\\[-1.00ex]
			Generic Information & GI & $\{$ $W_{t}$, $I^{i}$, $\Delta V_{t}^{i}$ $\}$\\[0.50ex]
			\hline \\[-1.0ex]
		\end{tabular}
	\end{adjustbox}
	\end{table} 
	
	\subsection{Statistical Learning\label{Sec:Method_StatLearn}}
	
	Gradient boosted classification and regression trees are used to understand the data and predict the test sets. This choice results from the favorable results of the method in \textcite{Krauss.2017}, as well as the fact that gradient boosted trees are among the models that most frequently win data mining competitions nowadays. 
	
	Boosting refers to a method proposed by \textcite{Schapire.1990}. It was introduced as an approach to combine many weak learners to one strong predictor, by fitting them not in parallel, but subsequently, while constantly reweighing the importance of explanatory variables. The final learner is built as an ensemble of the individually weak predictors. Fitting a single weak predictor is called a boosting iteration. After a single predictor is fit, it is used to predict the training data. Based on the residual errors produced, the training data is reweighed, by attaching weights to individual observations, so that the next learner focuses more on the incorrectly predicted observations of its predecessor. This kind of approach usually leads to an improvement in model accuracy over the unboosted method \parencite{Hastie.2017}. Applying the idea of boosting to classification problems, \textcite{Freund.1997} published the \textit{AdaBoost} algorithm, which gained a substantial amount of popularity both in research and in the field. Over time, the boosting algorithm has been subject to modifications and improvements which have led to a variety of implementations, most prominently the idea of gradient boosting, proposed by \textcite{Friedman.2002, Friedman.2001}. Gradient boosting fits weak learners to training data, as the classical boosting algorithm, but subsequently assumes an arbitrary loss-function to be minimized. At this moment gradient boosting fits predictors only to the residuals remaining after the first boosting iterations. Additionally, the gradient boosted model of \textcite{Friedman.2002} builds individual trees only on a random fraction of training observations, instead of the complete data, introducing randomness and subsequently improving the predictive power of the learner.
	
	The ``gbm" package in ``R" is a state-of-the-art implementation of gradient boosted trees, easy to use and sufficiently fast in computations. 
	
	In order to answer the question of the predictive value of different inputs, models are built on the subsets of the predictor sets $S = \{CR, SVI, GI\}$ (cf. Table~\ref{tbl:PredictorSets}). The empty set is excluded. This approach results in $2^{|S|}-1 = 7$ different subsets and models to be evaluated, summarized in Table~\ref{tbl:Models}. Component-wise subset creation is not considered, due to the exponential growth of the number of possible subsets. Computational feasibility can, consequentially, be ensured.
	
	\begin{table}[!htbp] 
	\caption[Summary of Models]{Summary of Models \newline The table lists the different subsets of predictors used for training the corresponding models. Components of the different predictor sets CR, SVI and GI can be found in Table~\ref{tbl:PredictorSets}.}
	\label{tbl:Models}
	\centering
	\footnotesize
	\begin{adjustbox}{max width=15.2cm}
		\renewcommand{\arraystretch}{0.8}
		\begin{tabular}{@{\extracolsep{5pt}}lc} 
			\\[-1.0ex]\hline 
			\hline \\[-1.0ex] 
			Model & Predictor Sets \\[+0.5ex]
			\hline \\[-1.0ex] 
			Model (CR) & \{ CR \}\\
			Model (SVI) & \{ SVI \}\\
			Model (GI) & \{ GI \}\\
			Model (CR, SVI) & \{ CR, SVI \}\\
			Model (CR, GI) & \{ CR, GI \}\\
			Model (SVI, GI) & \{ SVI, GI \}\\
			Model (CR, SVI, GI) & \{ CR, SVI, GI \}\\
			\hline \\[-1.0ex]
		\end{tabular}
	\end{adjustbox}
	\end{table} 
	
\begin{figure}[htbp] 
	\centering
	\footnotesize
	\includegraphics[width=0.5\textwidth]{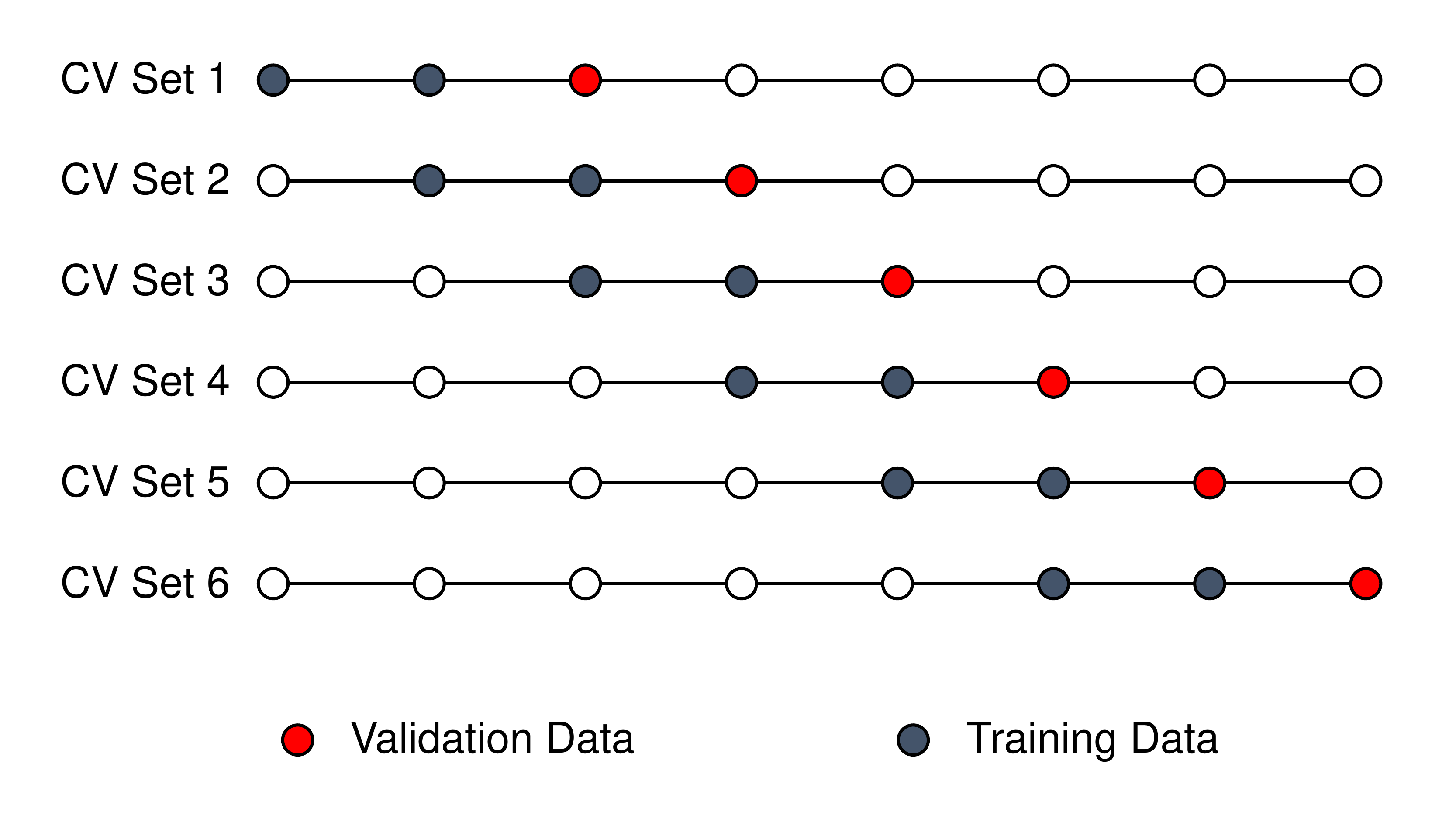}
	\captionsetup{justification=justified} 
	\caption[Implementation of the Cross Validation]{Implementation of the Cross Validation \newline Each dot represents approximately three months of data from the initial training set.} 
	\label{fig:CV_approach}
	\end{figure}

	Gradient boosting is prone to overfitting. To prevent overfitting, tuning parameters of each model on every training set are selected using cross-validation. Due to using past information as explanatory variables, the data exhibits a vital time series characteristic, where the chronological order of observations is important. In order to prevent models from picking up information originating chronologically after the validation set, k-fold cross-validation cannot be applied for parameter tuning. Differently, rolling cross-validation is applied, emulating real-world behavior and copying the idea of the subset creation, depicted in Figure~\ref{fig:Subsetting}. In consequence, each training set is divided into eight equally sized, chronologically coherent and non-overlapping folds. Models are validated by training them on two consecutive folds and predicting the next chronologically following fold. The approach is visualized in Figure~\ref{fig:CV_approach}, and follows \textcite{Callen.1996} and \textcite{Swanson.1997}, in evaluating models on rolling and equally sized training sets. The model with the highest average area under the receiver operating characteristic (ROC) over all validation folds is picked for prediction making on the test set. Opting for eight individual folds assures that every model is trained on a sufficient amount of data, which is about six months of observations. In consequence, each cross-validation fits models to six sets of training and validation data.
	
	Due to the high amount of data, model fitting is computationally a complex task. Hence parameter tuning only considers the interaction depth of the decision trees fitted during a single boosting iteration. Following \textcite{Hastie.2017}, interaction depths of $d \in \{4, 6, 8\}$ are considered in the cross-validation, in order to account for a sufficient amount of interactions between the explanatory variables. The number of trees fitted for each model, which is also the number of boosting iterations, is set to 500. This choice is higher than in the related work of \textcite{Krauss.2017}. Therefore, the learning rate, or otherwise referred to as shrinkage of the model, is analogously set lower to a value of 0.02.  Additionally, for each tree only half of the predictors are used, introducing randomness to the set of training observations. Additional parameters of the ``gbm" package are either not set or left at their default values. Tuning only one tuning parameter and setting others to relatively arbitrary values is an important choice for the sake of achieving computational feasibility. Consequentially, models will not represent global optima on the training sets. Overall results, however, should not be affected in their interpretability. 
	
	\subsubsection{Robustness Checking\label{Sec:Method_Robustness}}
	
	Due to the relatively arbitrary choice of two of the model's tuning parameters, an additional model is fit using a varying parametrization. Comparable results between the two iterations would suggest a higher general applicability. The general approach of subsetting and variable creation is not varied. For robustness checking, models with only 100 boosting iterations are fit to the training sets with a learning rate of 0.1. This approach differs from most of the present literature, as the number of trees is relatively low \parencite[cf.][]{Hastie.2017}. This low number of boosting iterations is similar to the approach used by \textcite{Krauss.2017}. Subsequently, the higher learning rate puts more emphasis on single boosting iterations and results in a faster learner. This parametrization was proven to be a powerful classifier in a similar context.
	
	\subsubsection{Prediction Making and Trading\label{Sec:Method_StatLearn_PredMaking}}
	
	The models try to predict whether or not a stock outperforms the daily median return of the index. The response of the gradient boosted tree on the test set is hereby interpreted as the probability to achieve outperformance. In order to test whether or not this information can be used for successful trading on financial markets, predictions on the test set are translated into a trading strategy: Every day the five stocks with the largest response are bought, whereas the five stocks with the lowest response value are sold short. Portfolios are equally weighted, such that the absolute sizes of long and short positions are the same. The application of this approach results in daily investment portfolios of 10 stocks for each of the models. On average, about 10\% of the index is traded each trading day.The idea of trading only top and bottom percentiles is motivated by \textcite{Huck.2009, Huck.2010}, in order to ignore observations with uncertainty about the directional movement of stock prices. The approach creates market neutral portfolios, as investors gain as much capital from short selling stocks as they spend on buying stocks. Net-investment at the beginning of a trading day is zero.
	
	\subsubsection{Statistical Learning Summary\label{Sec:Method_Summary}}
	
	In total, the cross-validation approach fits gradient boosted trees for each of the six CV sets for each of the seven different models (including different predictor sets) to each of the eleven particular time frames. Considering three levels of interaction depth results in (6 x 7 x 11 x 3) = 1,386 models for both the baseline approach and the robustness checking. On average, cross-validation training data contains 6,282 observations.
	
	After the cross-validation, a final gradient boosted tree is fit for every model on each of the initial training sets, resulting in (7 x 11) = 77 individual models for both the baseline approach and the robustness checking, with average training data containing 50,254 observations.
	
	Overall, the entire approach involved fitting 2,926 gradient boosted trees with a total of 877,800 individual boosting iterations. The large numbers and high complexity of the approach further underline the importance of simplifying the selection of tuning parameters.
		
	\section{Results\label{Sec:Results}}
	
	The results of the analysis cover two categories: (a) The overall quality of the gradient boosted models which is analyzed by each model's receiver operating characteristic and (b) the generated returns of the derived trading strategies on the test sets, covering mean returns and return distributions. Overall,  especially Model (CR, GI) and Model (CR, SVI, GI) perform better than the other. These two models have an average area under the receiver operating characteristics (AUC) of 0.566 and 0.567. The corresponding trading strategies generate average daily returns of 0.12\% per day compared to the overall index with 0.03\% per day.
	
\begin{figure}[htbp] 
		\centering
		\footnotesize
		\includegraphics[width=0.72\textwidth]{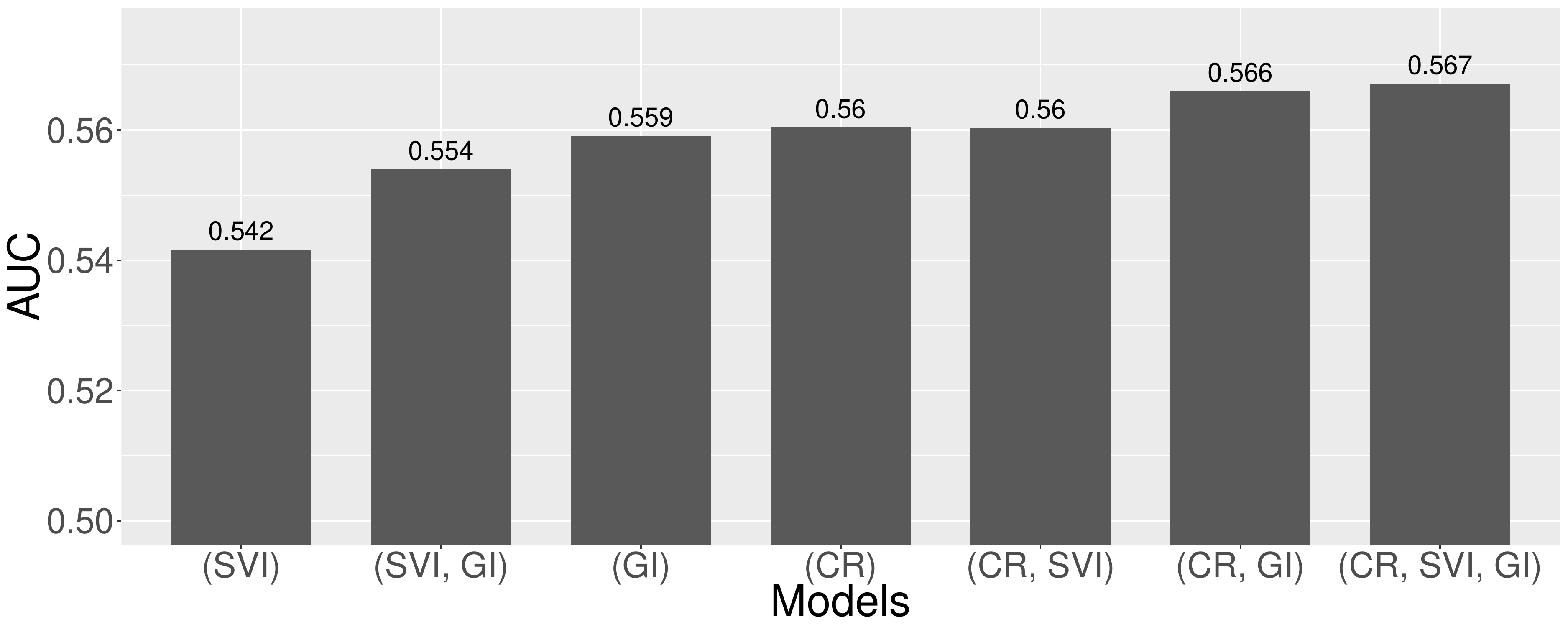}
		\captionsetup{justification=justified} 
		\caption[Predictive Quality of Models - Areas under the Receiver Operating Statistics]{Predictive Quality of Models - Areas under the Receiver Operating Statistics} 
	\label{fig:AUCs}
	\end{figure}

	
	All models fitted to the data show overall predictive capacity. Based on the daily average, all models obtain an AUC greater than 0.5, concluding that predictions are different from random guessing (cf. Figure~\ref{fig:AUCs}). Models that include cumulative returns (predictor set CR) obtain higher scores than the rest of the models, with Model (CR, SVI, GI) leading with an AUC of 0.567. The ranking of AUCs is in line with the returns generated by the specified trading strategy of trading ten stocks per day only, as further described below. Figure~\ref{fig:AverageAUCs} in the appendix displays average daily AUCs of all models over time. For all but Model (SVI) daily AUCs are downward trending with a statistically significant trend.
	
	Translated into trading strategies all models including cumulative returns (models (CR), (CR, SVI), (CR, GI) and (CR, SVI, GI)) create statistically significant average returns larger than 0, with t-statistics strictly larger than 2 ($H_0:$ ``Mean return equal to zero", critical value at the 5\% level: 1.9608). Additionally, models (CR, GI) and (CR, SVI, GI) proof to generate statistically significant average returns larger than the index with t-statistics strictly larger than 2 ($H_0:$ ``Mean returns are equal", critical value at the 5\% level: 1.9604). Vice versa, models without predictor set CR can not beat the index on average and perform worse. 
	
	Moreover, apart from Model (SVI), all other models create daily returns with a higher standard deviation than the index itself, in line with stated values of kurtosis of daily returns. All strategies create returns with an overall negative skewness, where Model (CR, SVI, GI) achieves the highest values with -0.0906, compared to Model (CR, SVI) with -0.7331, which has the lowest skewness within the different strategies.
	
	\begin{table}[!htbp] 
	\caption[Daily Return Statistics]{Daily Return Statistics \newline The table summarizes the characteristics of daily returns generated by the different sets of predictors compared to the general performance of the index.}
	\label{tbl:Results_MeanRet}
	\centering
	\footnotesize
	\begin{adjustbox}{max width=15.2cm}
		\renewcommand{\arraystretch}{0.8}
		\begin{tabular}{@{\extracolsep{5pt}}lcccccccc} 
			\\[-1.0ex]\hline 
			\hline \\[-1.0ex] 	
			& \multicolumn{6}{c}{Returns} \\		
			Statistic &         Model  &  Model  &  Model  &  Model  &  Model  &  Model  & Model   & S\&P 100\\
			& (CR) & (SVI) & (GI) & (CR, SVI) & (CR, GI) & (SVI, GI) & (CR, SVI, GI) &\\
			\hline \\[-1.0ex] 
			Minimum          &  -0.2750 &  -0.1751 &  -0.1990 &  -0.2816 &  -0.2716 &  -0.1851 &  -0.2896 &  -0.0878\\
			Quartile 1       &  -0.0063 &  -0.0048 &  -0.0063 &  -0.0059 &  -0.0061 &  -0.0056 &  -0.0061 &  -0.0039\\
			Median           &   0.0009 &   0.0001 &   0.0004 &   0.0007 &   0.0010 &   0.0005 &   0.0008 &   0.0006\\
			Arithmetic Mean  &   0.0007 &  -0.0001 &   0.0001 &   0.0010 &   0.0012 &   0.0002 &   0.0012 &   0.0003\\
			Geometric Mean   &   0.0005 &  -0.0002 &   0.0000 &   0.0008 &   0.0010 &   0.0001 &   0.0010 &   0.0002\\
			Quartile 3       &   0.0076 &   0.0047 &   0.0069 &   0.0076 &   0.0079 &   0.0058 &   0.0079 &   0.0053\\
			Maximum          &   0.1722 &   0.1728 &   0.2045 &   0.1621 &   0.1761 &   0.2270 &   0.2058 &   0.1124\\
			SE Mean          &   0.0003 &   0.0002 &   0.0003 &   0.0003 &   0.0004 &   0.0003 &   0.0004 &   0.0002\\
			LCL Mean (0.95)  &   0.0000 &  -0.0006 &  -0.0005 &   0.0003 &   0.0004 &  -0.0004 &   0.0005 &  -0.0002\\
			UCL Mean (0.95)  &   0.0014 &   0.0003 &   0.0008 &   0.0016 &   0.0019 &   0.0009 &   0.0019 &   0.0007\\
			Variance         &   0.0003 &   0.0001 &   0.0003 &   0.0003 &   0.0004 &   0.0003 &   0.0004 &   0.0001\\
			Stdev            &   0.0184 &   0.0119 &   0.0169 &   0.0180 &   0.0194 &   0.0167 &   0.0191 &   0.0122\\
			Skewness         &  -0.4946 &  -0.2813 &  -0.2393 &  -0.7331 &  -0.4554 &  -0.4413 &  -0.0906 &  -0.0500\\
			Kurtosis         &  34.2364 &  44.4771 &  25.0985 &  38.5705 &  29.4342 &  31.0572 &  36.9186 &  11.4246\\
			\hline \\[-1.0ex]
		\end{tabular}
	\end{adjustbox}
	\end{table} 
	
\begin{figure}[htbp] 
	\centering
	\footnotesize
	\includegraphics[width=0.8\textwidth]{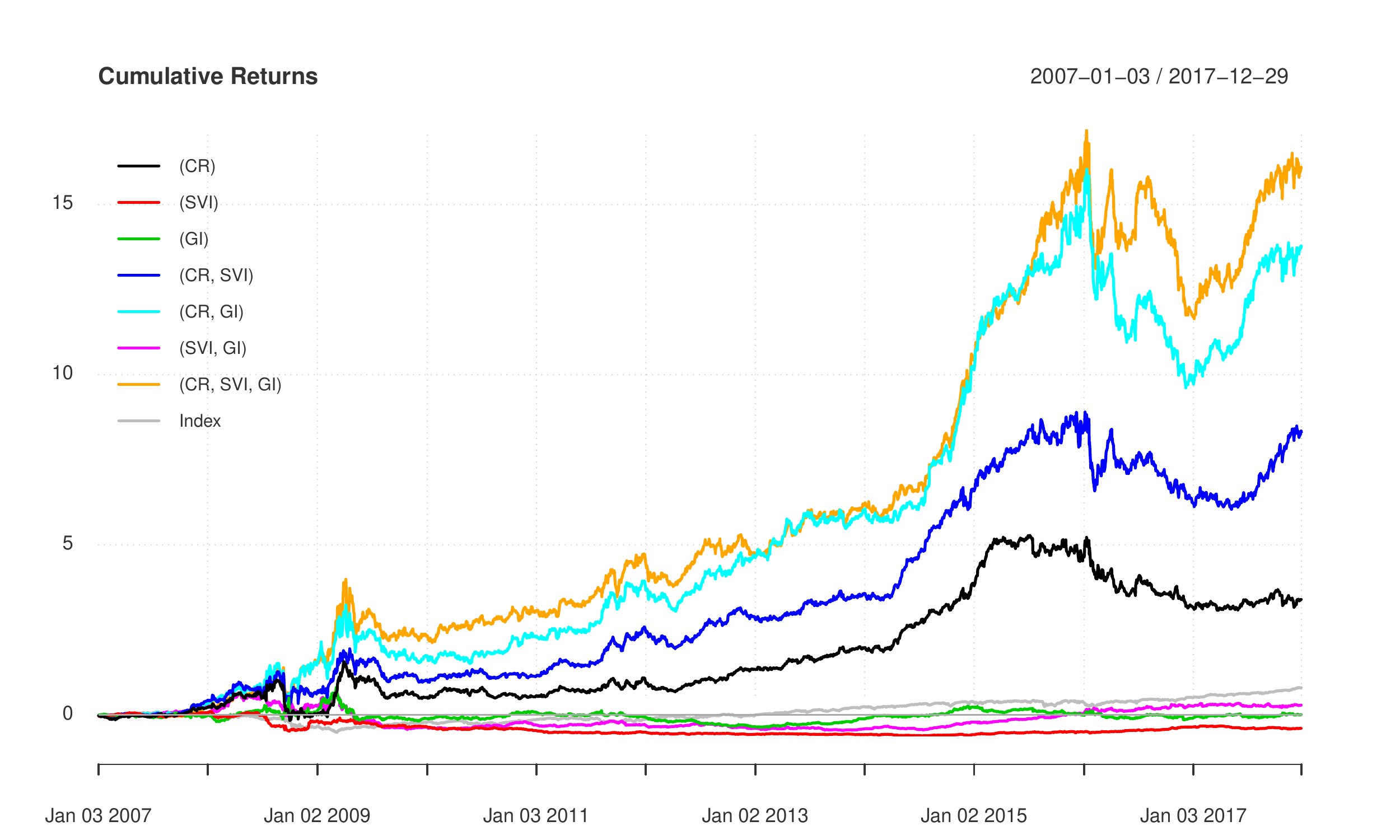}
	\captionsetup{justification=justified} 
	\caption[Cumulative Returns per Model]{Cumulative Returns per Model} 
	\label{fig:Cum_Returns}
	\end{figure}

\begin{figure}[htbp] 
	\centering
	\subfigure[2007 - 2009]{\label{SubP1}\includegraphics*[width=.45\textwidth]{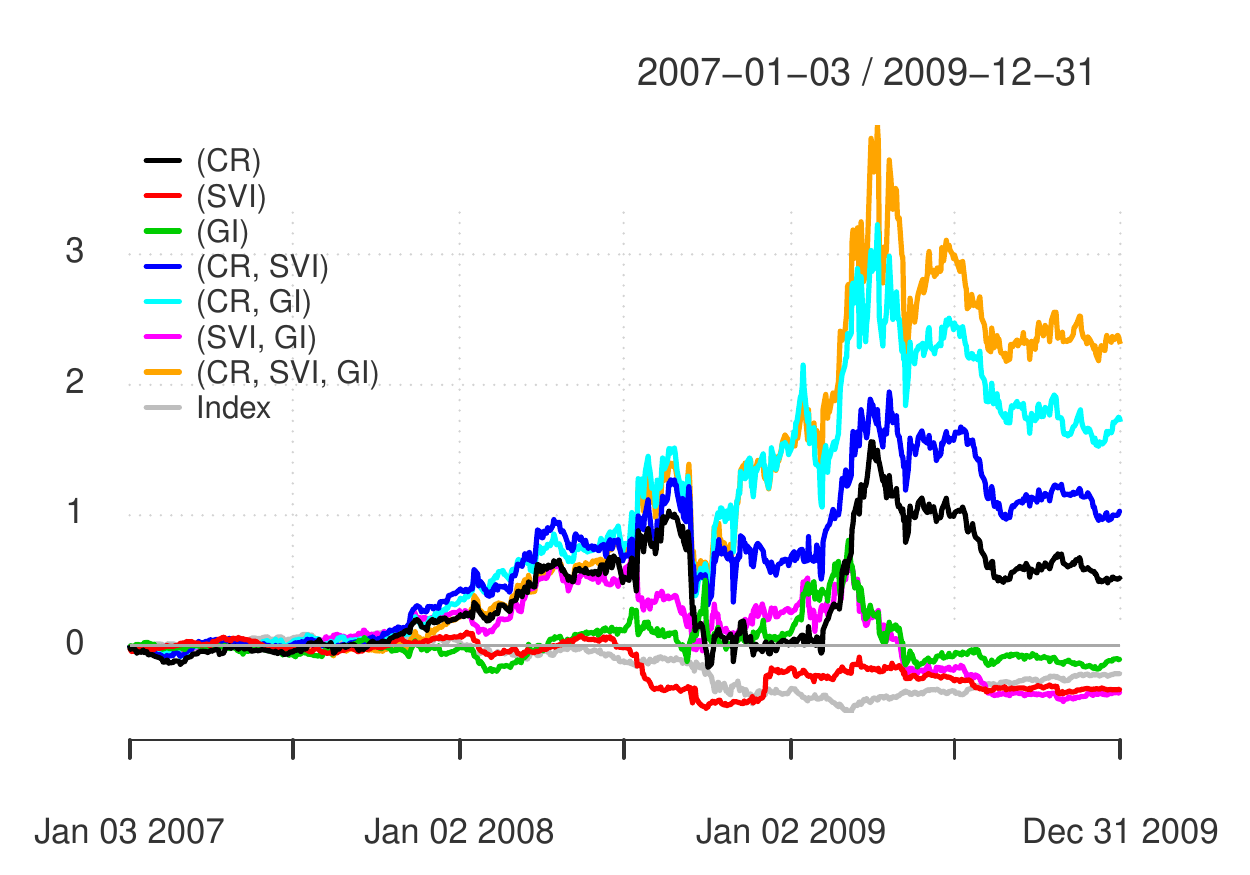}}
	\subfigure[2010 - 2012]{\label{SubP2}\includegraphics*[width=.45\textwidth]{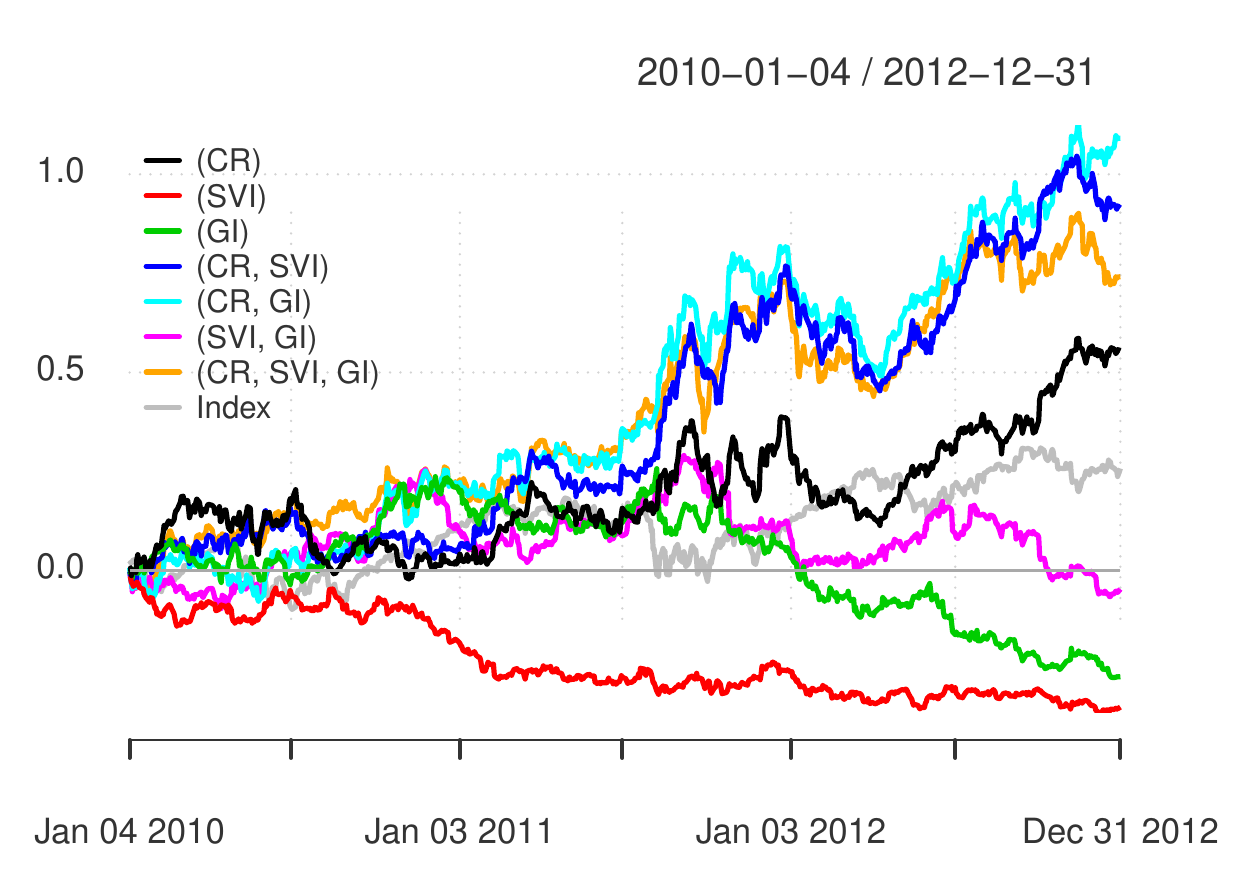}}
	\subfigure[2013 - 2015]{\label{SubP3}\includegraphics*[width=.45\textwidth]{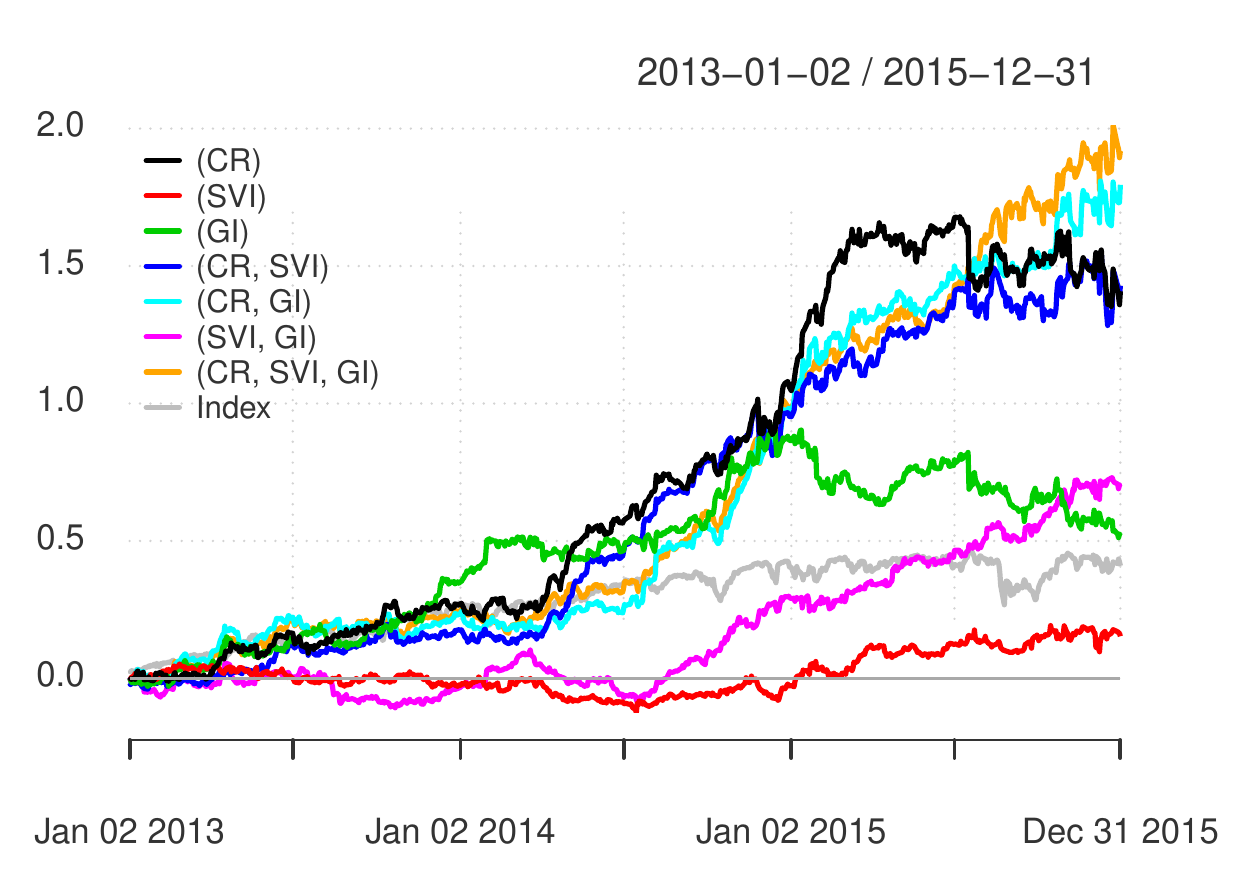}}
	\subfigure[2016 - 2017]{\label{SubP4}\includegraphics*[width=.45\textwidth]{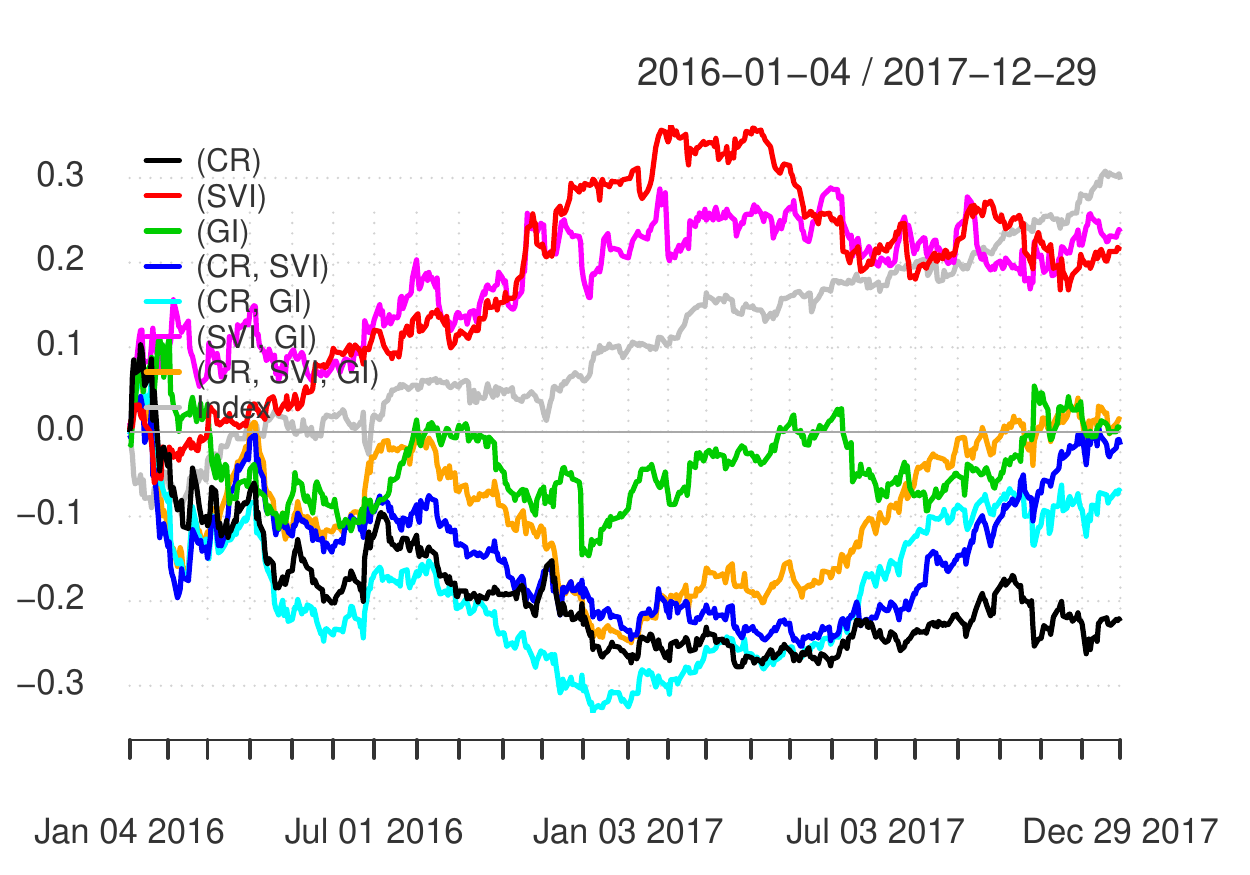}}
	\captionsetup{justification=justified} 
	\caption[Cumulative Returns on Subperiods]{Cumulative Returns on Subperiods} 
	\label{fig:SubPeriodRets}
	\end{figure}

	By plotting the generated cumulative returns of the trading strategies per model (cf. Figure~\ref{fig:Cum_Returns}) differences in performance over time become evident. This observation leads to a closer analysis of sub-periods. Four sub-periods are defined: Period 1 consists of the years 2007 to 2009, period 2 of the years 2010 to 2012, period 3 of the years 2013 to 2015 and period 4 of the years 2016 and 2017. Sub-period cumulative returns are depicted in Figure~\ref{fig:SubPeriodRets}. In the first three sub-periods results are comparable to what is stated for the overall performance of models on the entire time frame. Models using historical returns as predictors perform better than the index, other models perform comparably to the S\&P 100 or worse. Model (CR, SVI, GI) performs best in sub-period 1 and 3, whereas in sub-period 2 Model (CR, GI) is the overall best performer. Sub-period 4 provides a different picture, where strategies derived from models including predictor set (CR) perform the worst. The best performing strategy is derived from model (SVI, GI) closely followed by model (GI). Further, no strategy can perform better than the index over the time frame of sub-period 4.
	
	Examining strategy performance there is a difference in generated returns of stocks bought and stocks sold short, as shown in Figure~\ref{fig:Cum_Returns_long} and Figure~\ref{fig:Cum_Returns_short} in the appendix. Overall, the strategies perform better in identifying stocks to buy, than in identifying stocks to sell short. A more detailed summary of average returns generated by the long and short portfolios can be found in Table~\ref{tbl:Results_MeanRet_Long} and Table~\ref{tbl:Results_MeanRet_Short} in the appendix.
	
\begin{figure}[htbp] 
	\centering
	\subfigure[Model (CR)]{\label{Var1}\includegraphics*[width=.24\textwidth]{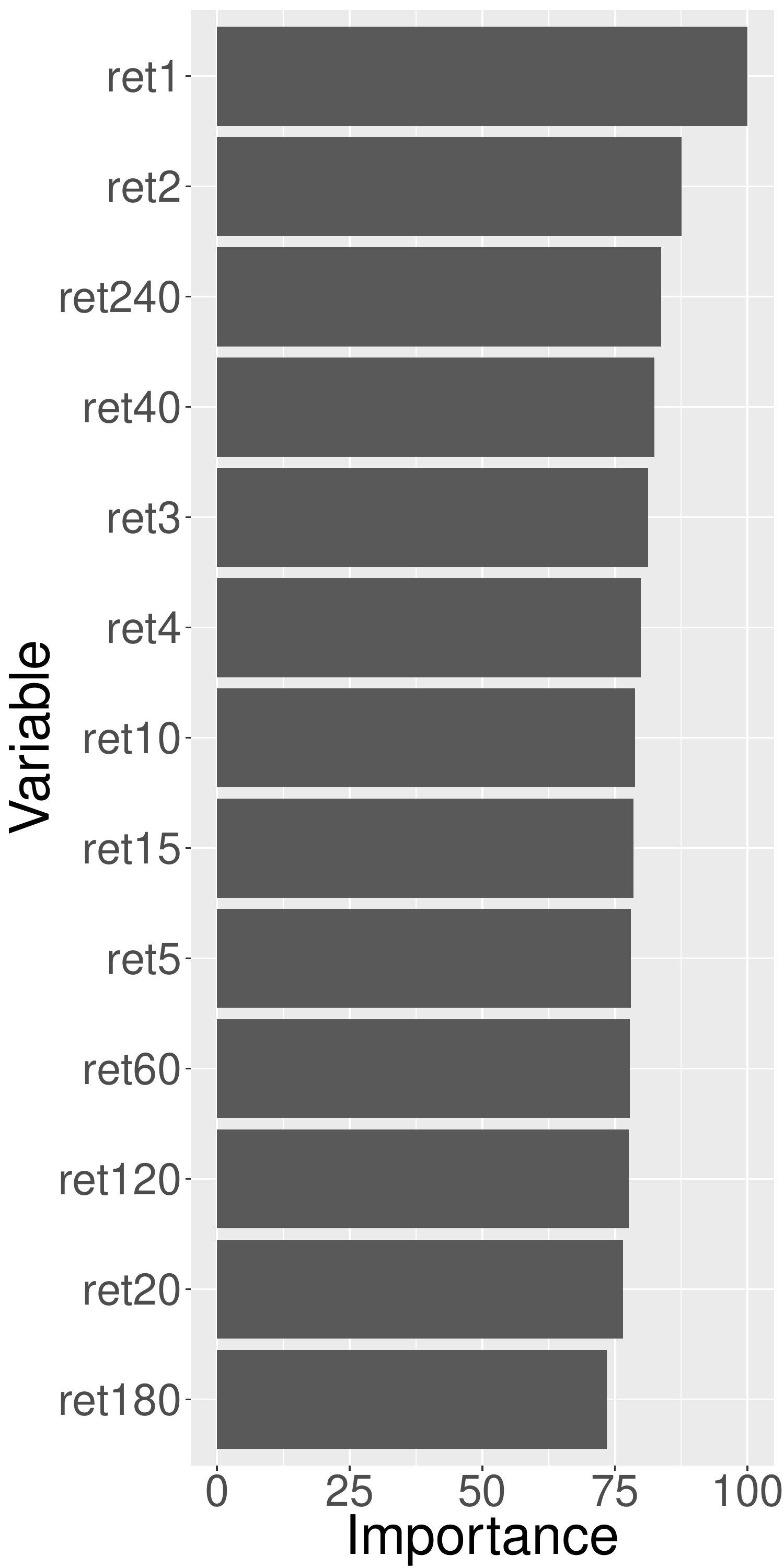}}
	\subfigure[Model (CR, SVI)]{\label{Var4}\includegraphics*[width=.24\textwidth]{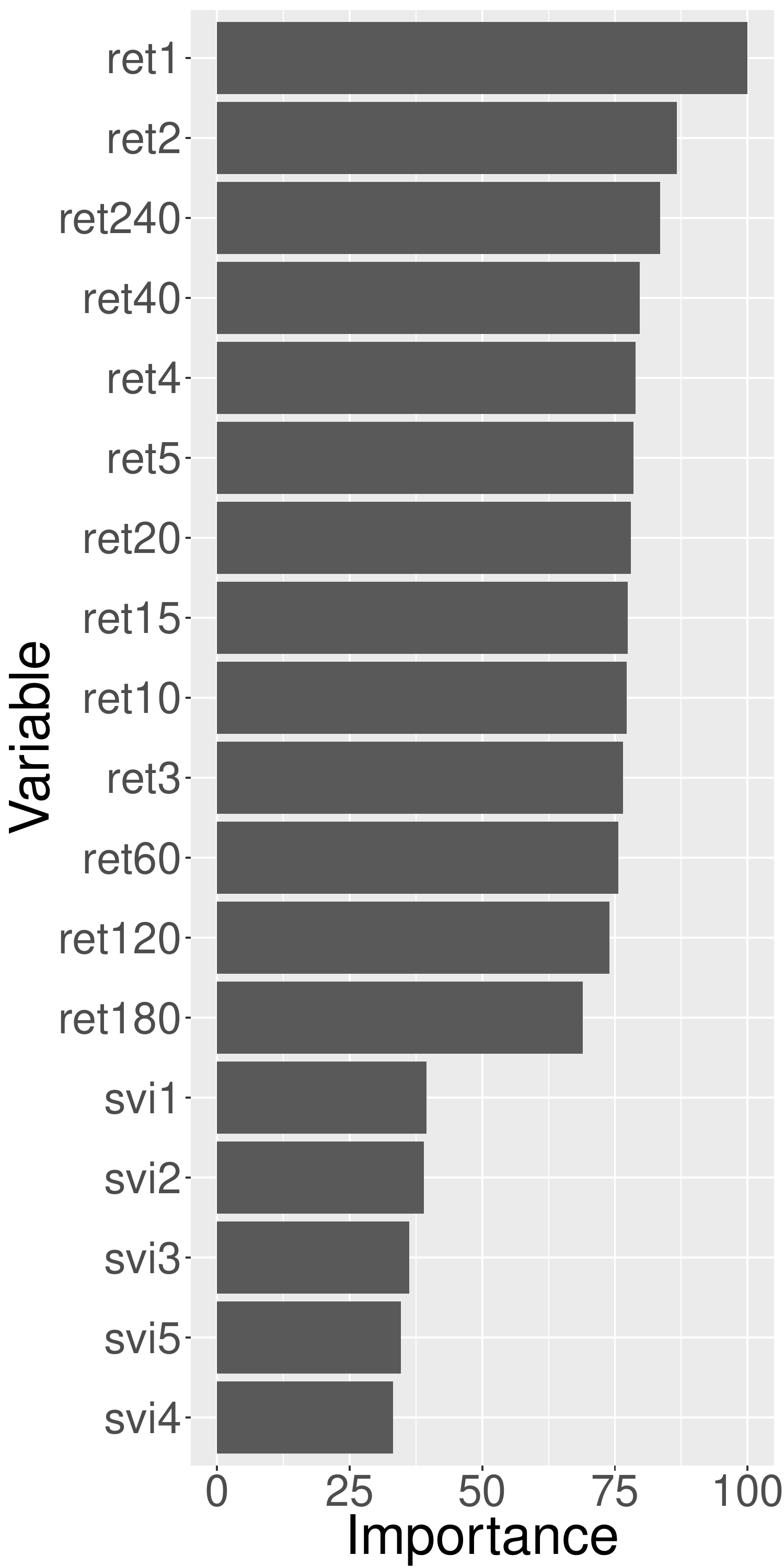}}
	\subfigure[Model (CR, GI)]{\label{Var5}\includegraphics*[width=.24\textwidth]{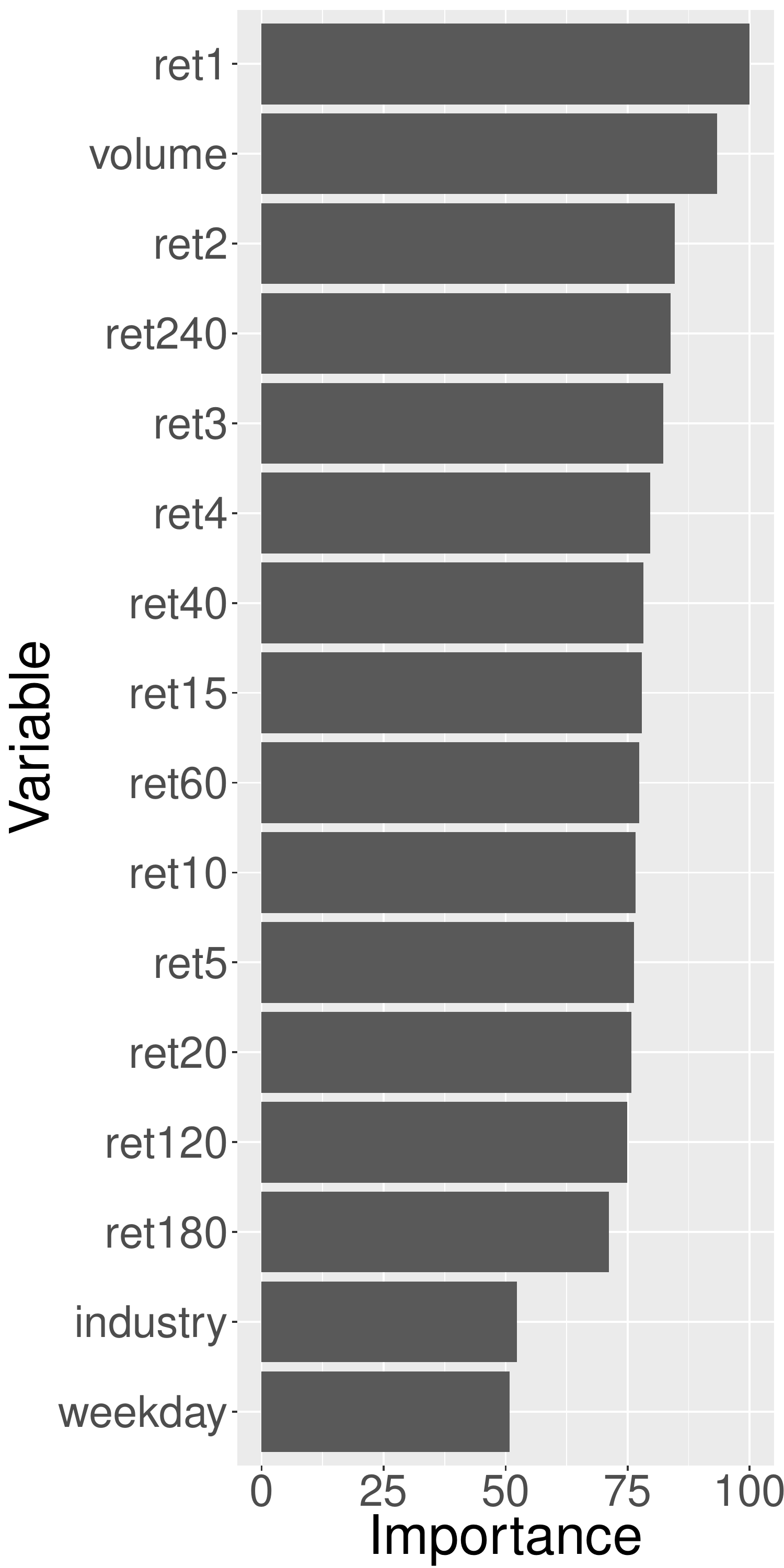}}
	\subfigure[Model (CR, SVI, GI)]{\label{Var7}\includegraphics*[width=.24\textwidth]{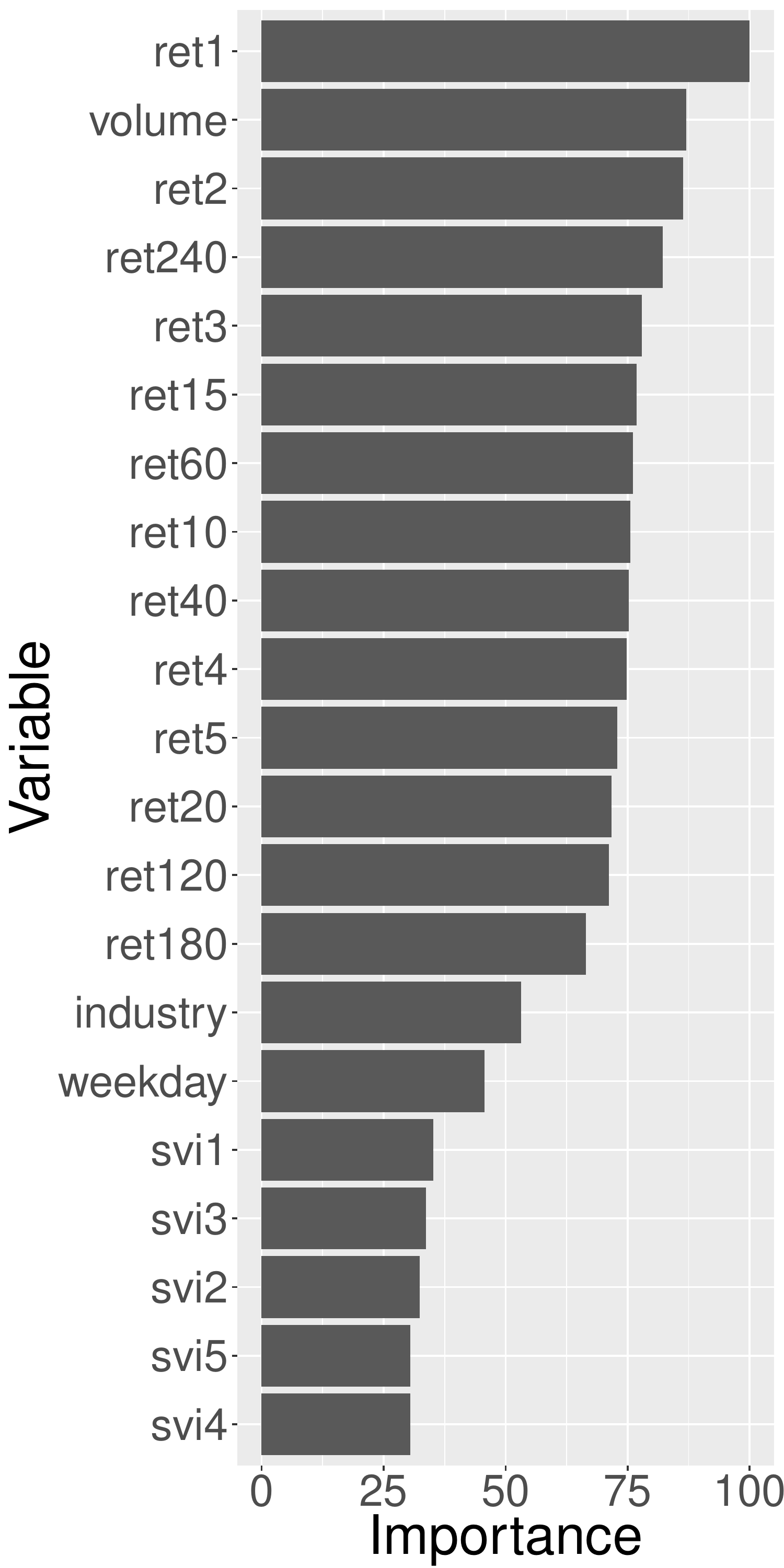}}
	\captionsetup{justification=justified} 
	\caption[Variable Importance per Model]{Variable Importance per Model} 
	\label{fig:VarImportance_1457}
	\end{figure}

	Finally, the relative variable importance of the models shows which of the explanatory variables have the highest influence on model and forecast accuracy. The variable importance thereby measures how often a variable is chosen for creating splits during the build process of the boosted decision tree, reweighed by the improvements this split had on the performance of the model on the training data \parencite{Elith.2008}. As eleven different boosted trees are fit for each of the seven models, variable importances are averaged and re-scaled such that the most critical variable obtains a score of 100. Figure~\ref{fig:VarImportance_1457} shows the average variable importance for all models including predictor set CR. Importance plots for the remaining models are found in Figure~\ref{fig:VarImportance_236} in the appendix. Generally, the historical returns of Set CR are the most important predictors followed by the generic predictors of Set GI and lastly the search volume indices of Set SVI. However, it is noteworthy that the change in trading volume, if included in the model, is the second most important variable. Overall, all of the variables have a non-zero influence. With both the historical returns and SVIs, observations from the more recent past make more important predictors. Most importantly, the last-day return functions as the most important predictor in all the models including predictor set CR.
	
	Analysing the test for overall robustness, i.e. training gradient boosted models with 100 boosting iterations and a shrinkage parameter of 0.1, a pattern comparable to the baseline analysis emerges, i.e. training gradient boosted models with 500 boosting iterations and a shrinkage parameter of 0.02. Models including historical returns classify more accurately and result in trading strategies outperforming those models without the bespoken explanatory variables. Again, models including predictor set CR create performance in excess of the S\&P 100, whereas the remaining models cannot beat the index. The results are visualized in Figure~\ref{fig:Cum_Returns_Robustness} in the appendix. However, some variations in performances are noteworthy: (i) Overall, CR models, even though still performing better than the others, create less cumulative return over the time frame of the analysis. Model (CR, SVI, GI), for example, creates more than 1500\% of return with the baseline parametrization but only just under 700\% with the robustness parametrization. Nonetheless, it is still the best performing model. (ii) While Model (CR, GI) performs better than Model (CR, SVI) in the baseline setting, the situation reverses after altering tuning parameters. (iii) Model (SVI) and Model (GI) create overall negative returns at the end of the time frame with the baseline parametrization, but create positive performance at the end of the period in the robustness checking. Thus, all models in this setting create positive returns.
	
	\section{Discussion\label{Sec:Discus}}
	
	The following section discusses insights from the results of the analysis, as well as evaluated contributions to existing research:
	
	\subsection{General Results}
	
	Overall forecasting results and the performance of trading strategies provide clear evidence for the predictive capacity of all models and especially predictor sets used for this study. Even though three of the seven considered models implement trading strategies inferior to the general market return, all models have an average AUC greater than 0.5 which is an important quality for any predictive model. Especially four models implement strategies creating substantial amounts of cumulative return over time. All of them include historical returns as explanatory variables. Especially the (CR, SVI, GI) model, using all of the available predictor sets, creates average daily returns four times the size of the underlying index. The approach would have created a total of more than 1500\% of return over 11 years of trading.
	
	Models not including historical returns, however, provide a less clear picture. Between 2007 and 2015 these models predict worse than other models and result in trading returns similar to or smaller than the index. Only in 2016, these models seem to forecast more accurately than models including historical returns.
	
	Moreover, results suggest a superior performance of models using more than one set of predictors. The favourable results of model (CR, SVI, GI) suggest that all predictor sets contain valuable information. Also, the predictor sets appear to have little correlation, as adding a set usually increases model performance.

	\subsubsection{Value of Enriched Datasets}
	
	The results suggest that including data on internet search engine usage as well as generic information increases the overall performance of models predicting future stock returns. Variable importances show that the predictors from all predictor sets, i.e. historical cumulative returns, search volume indices and generic information, have non-zero influence (cf. Figure~\ref{fig:VarImportance_1457} and Figure~\ref{fig:VarImportance_236}). While it is surprising that such simple information as classifying trading days and industries lead to increased performances, particular focus should be laid upon the value of data on search engine usage. The increase in model performance is supporting the assumption that there is a relationship between internet user behavior and future action of investors.
	
	Focusing especially on the (SVI) model, which solely uses predictors derived from the Google Trends database, the predictor shows varying performance. In the first two sub-periods of the analysis the investment strategy derived from the predictors generates negative returns altogether. Only in 2014 returns appear to be positive on average. More strikingly, the (SVI) model implements the second most profitable strategy in the last sub-period from 2016 to 2017. The strategy is second to only the (SVI, GI) model, using an ensemble of search engine data and generic indicators.
	
	Reasons for this improvement might be the increase of the market share of Google over the considered time frame, as seen in \textcite{StatCounter.2019}. Additionally, Google added various features to their search engine over time, such as the Knowledge Graph \parencite{Singhal.2012}, summarizing information on the level of the search engine, or the financial database Google Finance\footnote{As of 09/04/2019 Google Finance is accessible via \url{https://www.google.com/finance}.}, reducing search costs of users and potentially increasing usage of the engine for research purposes. This claim is supported by the change of data composition depicted in Section~\ref{Sec:Data and Proc}, especially Figure~\ref{fig:Mean_SVI} and Figure~\ref{fig:0_values_SVI}, as the average SVI is increasing and the relative share of missing values decreasing over time.
	
	\subsection{Important Contributions}
	
	\subsubsection{On Financial Forecasting}
	
	The performance of ensemble models demonstrates that both generic information as well as internet user behaviour provide valuable insights about future returns of stock prices. Empirical returns of the strategies implemented offer a new view on how this information can be exploited to create consistent positive returns. This paper thereby adds to \textcite{Choi.2012}, \textcite{Ettredge.2005} and \textcite{Preis.2012} in so far as stock returns are yet another field that can partly be predicted by internet search data.
	
	The study extends the work of \textcite{Bordino.2012} on predicting stock market trading volumes. Results are providing evidence that internet user behavior can not only qualitatively predict whether or not trading volumes are going to increase or decrease. Especially in combination with other predictors and in a multivariate approach it shows the ability to quantify the directional performance of stock prices.
	
	\subsubsection{On Market Efficiency}
	
	The higher returns of the ensemble models further challenge the proposition of weak form efficient markets. According to \textcite{Malkiel.1970}, publicly available historical data should contain no valuable information for profitable investments into the market. The results of this study are opposing this hypothesis and are further in line with \textcite{Grossman.1980}, who claim that markets cannot be informationally efficient if information aggregation and processing are costly: 
	
	The high complexity of the analysis suggests that information processing is costly for researchers and investors. However, diminishing returns of all models over time suggest that the wide availability of processing techniques and computation power lowered the costs of processing substantially. A claim that is also made by \textcite{Krauss.2017}. Moreover, model performances over time show that not every source of information is exploited to the same extend. While models on financial data and generic information are performing worse over time, the AUC of the (SVI) model does not decrease. Analogously, the trading strategies derived from (SVI) and (SVI, GI) outperform the other strategies in the sub-period from 2016 to 2017. This observation and the high amount of time needed to download search engine data emphasize that in recent times the costs of data aggregation outweigh the costs of data processing. 
	
	\subsubsection{On Statistical Arbitrage}
	
	Especially the results from the (CR) model, only including historical cumulative returns as explanatory variables, are adding to the results of \textcite{Krauss.2017}. Their approach to statistical arbitrage in stock indices proves to be successful in the S\&P 100 as well. Although the index is a true subset of the S\&P 500, which was used for their study, the S\&P 100 varies significantly in essential characteristics, most importantly the number of constituents and average market capitalization of companies. With only a fifth of the components the index provides fewer input data for model training on companies that are well covered by analysts. This circumstance may partly explain why the performance of thee (CR) model in this study is smaller than the performance of the models considered by \textcite{Krauss.2017}. While the general approach is supported, further generalizations are dubious. The method still needs to be tested on stocks with less liquidity and analyst coverage.
	
	Further, the use of alternative data sources adds to the field of statistical arbitrage, both in academia and in the field. Internet user data and especially search engine query volumes might add a valuable, little exploited resource for predicting stock performances and other financial key figures. This is especially supported by the development of daily receiver operating characteristics of the different models. While all other models show a significant downward trend in daily AUCs, the (SVI) model, only using Google Trends data, produces less volatile model performances and exhibits no statistically significant downtrend.
	
	\subsection{Conduct Evaluation and Model Setup}
	
	Albeit the favourable results and the quality of predictive models, the general approach of the statistical analysis needs critical reflection.
	
	\subsubsection{Input Data}
	
	While data on historical returns, trading volumes and industry classifications is widely used in academia and the field, the usage of internet user behavior and especially the Google Trends database is little dealt with in the existing literature. This paper assumes that future investors interact with the internet before they trade stocks on the market. The relative search volumes monitor this interaction for individual pairs of ticker symbols and stocks. While results provide arguments that ticker symbols can be used for this purpose, it is noteworthy that the choice of an adequate search term is crucial for the success of the strategy. Beyond that, ticker symbols might not be the ideal choice for this paper's purpose. Investors might alternatively use the clear name of a company or might even search for relatively unrelated search terms, such as financial databases, industries, or even people, such as members of boards of directors. Examining an ideal set of considered search terms is out of scope for the extent of this paper and left open for future research. Possible approaches include varying search terms or including sets of multiple search terms.
	
	Besides that, the Google Trends database itself imposes limitations to model quality. As Google calculates SVI from random samples of its historical search data, time series vary over different downloads \parencite{Preis.2013}. Consistency could possibly be improved by multiple downloads of the same SVI and averaging. Doing so might additionally reduce the high amounts of missing data in the first years of the empirical study. The approach was not followed for the analysis of this paper due to the high number of downloads\footnote{Following the approach of creating consistent SVI time series from Section~\ref{Sec:Data and Proc_Data_Internet} leads to downloading about 20,000 individual time series from the Google Trends database. In order to average results several tens of thousands of downloads would have been needed. Next to download speed from the database, mainly quota limits restricted this study from following the formulation.}.
	
	\subsubsection{Feature Creation}
	
	The creation of explanatory variables from input data might be one of the most critical tasks to enable statistical learning and prediction making. While the way of creating categorical features is hardly questionable and straight forward, creating quantitative variables is less obvious in the case of this study. Creating features from historical returns is motivated by \textcite{Takeuchi.2013} and already proven successful in other pieces of academic research. In contrast, the transformation of internet search volumes is novel and might be up for further enhancements. Their non-zero influence and capability to enhance the predictive quality of models indicate that the current transformation is meaningful. A different transformation might nonetheless further increase model performances.
	
	\subsubsection{Model and Parameter Tuning}
	
	Concerning the purpose of this work, the choice of an ideal statistical learner as well as finding a globally optimal model is of secondary interest. Overall results might consequentially be subject to improvement by varying the underlying statistical model. Gradient boosting proves to be one of the most potent statistical learners. Still, \textcite{Hastie.2017} point out that there is not one single solution to understanding data best. The quality of models varies over different fields of application, as seen in \textcite{Kim.2018}. 
	
	Supplementary, the approach to tune tuning parameters of the boosted models (cf. Section~\ref{Sec:Method_StatLearn}) can likely be enhanced. First, only one parameter is considered. Ideally, all tuning parameters of the boosted model would be assessed. Second, the length of training and test sets is mostly arbitrary and only motivated by the approach of \textcite{Krauss.2017}. These variables remain the target of optimization as well. Third, indicators are grouped to predictor sets. Feature selection techniques could be used to find the most suitable model. Finally, models are evaluated based on their receiver operating characteristic. This operation optimizes the overall predictive quality of the model, though it might not be the correct metric to optimize returns of the trading strategy. To find the ideal model for strategy implementation models needed to be evaluated by strategy performance during cross-validation rather than the ROC.
	
	However, the approach followed for this analysis is successfully balancing computational feasibility and performance. All of the above enhancements would result in increasing complexity, making the process of finding and fitting models extremely time and cost intensive. 
	
	\subsubsection{Investment Strategy}
	
	Just as the training of the models, the investment strategy is subject to further improvements. Compared to \textcite{Krauss.2017}, the presented approach does not consider different sizes of trading portfolios (cf. Section~\ref{Sec:Method_StatLearn_PredMaking}). Their result advice that the overall ranking of performances does not change over differently sized trading portfolios. Still, the possibility to optimize this parameter should be noted. Again, this would increase the computational complexity of the approach.
	
	\subsubsection{Real-World Applicability}
	
	Real-world applicability of the approach cannot be judged based on the results. As a feasibility study, the analysis does not focus on financial performance indicators, the associated risk of the trading strategies or transaction costs. These figures need to be assessed individually, to be able to judge, whether the strategy is suitable to be implemented.
	
	\section{Conclusion\label{Sec:Concl}}
	
	This paper implemented a statistical arbitrage strategy, predicting directional stock price movements on the S\&P 100 index between 2005 and 2017. Models have been trained using historical stock returns, data on internet user behavior as well as complementary generic information in order to evaluate the data's predictive capacities. The approach attempted to test the weak form efficiency of financial capital markets, as well as the relationship between internet user behavior and stock price movements. These core elements create relevance of the work for both ongoing academic research and applications in the financial industry. Data included time series from the Google Trends database, a data source little exploited in academic literature to the current date, harvesting information about trillions of queries each year on the world's largest internet search engine \parencite{Sullivan.2016}. Models have been built using gradient boosted decision trees, a state-of-the-art statistical learning method, enabling to limit assumptions about the structure of the data and relationships between predictors. 
	
	Results show that this approach indeed trained models that successfully classified day-ahead returns of stocks. Receiver operating characteristics of all models indicate that either of the datasets contained valuable information for investing into stocks of the index, confirming the initial hypothesis of this work. Especially models using ensembles of the different data sources showed increased prediction quality. When using all of the available data, the approach managed to create a cumulative historical return before transaction costs of more than 1500\% between 2007 and 2017 and on average an area under the receiver operating characteristic of 56.7\%. On top of that, empirical results suggest that internet user behavior is a little exploited resource for investment decision making, as the historical model quality is not diminishing over time, in contrast to models using financial input data. The overall results imply important contributions to the existing literature:
	
	First, the results provide guidance for the usage of internet search data for financial modeling and forecasting. There is evidence for a relationship between future stock returns and the current behavior of internet users. The inclusion of historical data from the Google Trends database successfully increased model performances in the empirical study. However, this paper does not claim to have found a universal answer to the structure of this dependence. Moreover, the results suggest multivariate relationships between the performances of stocks and different, mutually uncorrelated sources of data. As the statistical analysis focused on the general feasibility of using alternative data sources, global model optimization had only secondary importance. The question of how to ideally transform input data and train models is therefore left open for future research.
	
	Second, the research adds to \textcite{Krauss.2017}, who provided the general framework of the analysis. Empirical results support their main hypotheses. Most importantly, results show that data-driven statistical arbitrage strategies can be applied to medium-sized stock indices and create returns in highly liquid and well-covered environments. Additionally, average receiver operating characteristics support their postulate of diminishing returns of models using historical returns. 
	
	Third, the high returns in excess of the market further question the weak form efficiency of modern financial capital markets, although results show that markets become more efficient over time, most likely due to the increasing availability of data processing equipment and techniques. The consistent predictive performance of internet data propounds further that the cost of comprehensive data aggregation increases relative to the cost of data processing, as novel sets of input data appear to be little exploited. The relationship between costs and value of aggregating relevant internet usage related information is left open for future research.
	
	Overall, this paper aims to have provided practical insights into using internet user behavior in the context of quantitative and data-driven financial markets investing. While the close-up financial evaluation of the strategy remains open for further evaluation, the high empirical returns suggest general feasibility of the approach and auspicious research. With the increasing technologization of the everyday life and professional industries the relevance of internet usage related data is likely to further increase in importance for accurate modeling and forecasting economic indicators and events.

\newpage
\addcontentsline{toc}{section}{References} \nocite{*}
\printbibliography 
\newpage
\setcounter{table}{0}
\renewcommand{\thetable}{A\arabic{table}}
\setcounter{figure}{0}
\renewcommand{\thefigure}{A\arabic{figure}}
\begin{appendix}
	\section{Appendix}
	
	\subsection{Tables}
	
	\begin{table}[!htbp] 
		\caption[Weekday Overview S\&P 100]{Weekday Overview S\&P 100 \newline The table summarizes daily mean returns by different weekdays.}
		\label{tbl:WeekdayOverview}
		\centering
		\footnotesize
		\begin{adjustbox}{max width=15.2cm}
			\renewcommand{\arraystretch}{0.8}
			\begin{tabular}{@{\extracolsep{5pt}}lcc} 
				\\[-1.0ex]\hline 
				\hline \\[-1.0ex] 
				Weekday &  Mean Return [\%] & Standard Deviation [\%]\\
				\hline \\[-1.0ex] 
				Monday & -0.007\% & 2.3\\
				Tuesday & 0.104\% & 2.2\\
				Wednesday & 0.067\% & 2.1\\
				Thursday & 0.066\% & 2.2\\
				Friday & 0.046\% & 2.0\\
				\hline \\[-1.0ex]
			\end{tabular}
		\end{adjustbox}
	\end{table}
	
	\begin{table}[!htbp] 
		\caption[Daily Return Statistics - Long only]{Daily Return Statistics - Long only \newline The table summarizes the characteristics of daily returns generated by the different sets of predictors and investing the five stocks with the highest probability to outperform the daily median of index returns. Values are compared to the general performance of the index.}
		\label{tbl:Results_MeanRet_Long}
		\centering
		\footnotesize
		\begin{adjustbox}{max width=15.2cm}
			\renewcommand{\arraystretch}{0.8}
			\begin{tabular}{@{\extracolsep{5pt}}lcccccccc} 
				\\[-1.0ex]\hline 
				\hline \\[-1.0ex] 	
				& \multicolumn{6}{c}{Returns} \\		
				Statistic &         Model  &  Model  &  Model  &  Model  &  Model  &  Model  & Model   & S\&P 100\\
				& (CR) & (SVI) & (GI) & (CR, SVI) & (CR, GI) & (SVI, GI) & (CR, SVI, GI) &\\
				\hline \\[-1.0ex] 
				Minimum           &   -0.3324 &  -0.2179 &  -0.1342 &  -0.3394 &  -0.3394 &  -0.2226 &  -0.3474 &  -0.0878\\
				Quartile 1        &   -0.0062 &  -0.0059 &  -0.0057 &  -0.0060 &  -0.0059 &  -0.0055 &  -0.0057 &  -0.0039\\
				Median            &    0.0010 &   0.0004 &   0.0009 &   0.0011 &   0.0009 &   0.0007 &   0.0009 &   0.0006\\
				Arithmetic Mean   &    0.0007 &   0.0003 &   0.0005 &   0.0008 &   0.0008 &   0.0004 &   0.0009 &   0.0003\\
				Geometric Mean    &    0.0004 &   0.0002 &   0.0004 &   0.0006 &   0.0006 &   0.0003 &   0.0007 &   0.0002\\
				Quartile 3        &    0.0081 &   0.0072 &   0.0071 &   0.0078 &   0.0080 &   0.0069 &   0.0081 &   0.0053\\
				Maximum           &    0.2211 &   0.2041 &   0.1461 &   0.1934 &   0.1970 &   0.1541 &   0.2229 &   0.1124\\
				SE Mean           &    0.0004 &   0.0003 &   0.0003 &   0.0004 &   0.0004 &   0.0003 &   0.0004  &  0.0002\\
				LCL Mean (0.95)   &   -0.0002 &  -0.0003 &   0.0000 &   0.0000 &   0.0000 &  -0.0001 &   0.0001 &  -0.0002\\
				UCL Mean (0.95)   &    0.0015 &   0.0010 &   0.0011 &   0.0016 &   0.0016 &   0.0010 &   0.0016 &   0.0007\\
				Variance          &    0.0005 &   0.0003 &   0.0002 &   0.0005 &   0.0004 &   0.0002 &   0.0004 &   0.0001\\
				Stdev             &    0.0222 &   0.0164 &   0.0153 &   0.0217 &   0.0210 &   0.0149 &   0.0207 &   0.0122\\
				Skewness          &   -0.9068 &  -0.0736 &   0.1039 &  -0.6180 &  -0.8407 &  -0.8572 &  -0.6939  & -0.0500\\
				Kurtosis          &   42.4654 &  31.2316 &  15.3632 &  37.7960 &  41.0555 &  28.1783 &  47.3824 &  11.4246\\
				\hline \\[-1.0ex]
			\end{tabular}
		\end{adjustbox}
	\end{table} 
	
	\begin{table}[!htbp] 
		\caption[Daily Return Statistics - Short only]{Daily Return Statistics - Short only \newline The table summarizes the characteristics of daily returns generated by the different sets of predictors and short selling the five stocks with the lowest probability to outperform the daily median of index returns. Values are compared to the general performance of the index.}
		\label{tbl:Results_MeanRet_Short}
		\centering
		\footnotesize
		\begin{adjustbox}{max width=15.2cm}
			\renewcommand{\arraystretch}{0.8}
			\begin{tabular}{@{\extracolsep{5pt}}lcccccccc} 
				\\[-1.0ex]\hline 
				\hline \\[-1.0ex] 	
				& \multicolumn{6}{c}{Returns} \\		
				Statistic &         Model  &  Model  &  Model  &  Model  &  Model  &  Model  & Model   & S\&P 100\\
				& (CR) & (SVI) & (GI) & (CR, SVI) & (CR, GI) & (SVI, GI) & (CR, SVI, GI) &\\
				\hline \\[-1.0ex] 
				Minimum            &  -0.1394&   -0.1200&   -0.2315&   -0.1400&   -0.1588&   -0.2089&   -0.1646&   -0.0878\\
				Quartile 1         &  -0.0085 &  -0.0067 &  -0.0080 &  -0.0083 &  -0.0084 &  -0.0073 &  -0.0083 &  -0.0039\\
				Median             &  -0.0001  & -0.0008  & -0.0003  &  0.0000  &  0.0001   &-0.0005  &  0.0000  &  0.0006\\
				Arithmetic Mean    &   0.0000   &-0.0005   &-0.0004   & 0.0001   & 0.0004&   -0.0002   & 0.0003   & 0.0003\\
				Geometric Mean     &  -0.0002&   -0.0006&   -0.0006   &-0.0001    &0.0001 &  -0.0004    &0.0001    &0.0002\\
				Quartile 3         &   0.0077 &   0.0054 &   0.0072&    0.0076&    0.0081  &  0.0068&    0.0081&    0.0053\\
				Maximum            &   0.1569  &  0.1361  &  0.3084 &   0.1545 &   0.2182   & 0.3007 &   0.1857 &   0.1124\\
				SE Mean            &   0.0004   & 0.0003   & 0.0004  &  0.0004  &  0.0004    &0.0004  &  0.0004  &  0.0002\\
				LCL Mean (0.95)    &  -0.0007&   -0.0010   &-0.0012   &-0.0006   &-0.0004&   -0.0010   &-0.0005   &-0.0002\\
				UCL Mean (0.95)    &   0.0008 &   0.0001&    0.0004    &0.0009&    0.0012 &   0.0006&    0.0011&    0.0007\\
				Variance           &   0.0004  &  0.0002 &   0.0005&    0.0004 &   0.0005  &  0.0005 &   0.0005 &   0.0001\\
				Stdev              &   0.0211   & 0.0151  &  0.0215 &   0.0207  &  0.0220   & 0.0212  &  0.0216  &  0.0122\\
				Skewness           &   0.4227    &0.3330   & 0.1337  &  0.3987   & 0.4447    &0.6432   & 0.3856   &-0.0500\\
				Kurtosis           &  10.5405&   14.0284   &31.1298   &10.3403   &14.0592   &28.3926   &13.3903   &11.4246\\
				\hline \\[-1.0ex]
			\end{tabular}
		\end{adjustbox}
	\end{table} 
	
	\newpage
	
	\subsection{Figures}
	
	\begin{figure}[htbp] 
		\centering
		\subfigure[Unadjusted Daily SVI vs. True Montly SVI]{\label{Unadjusted}\includegraphics*[width=.44\textwidth]{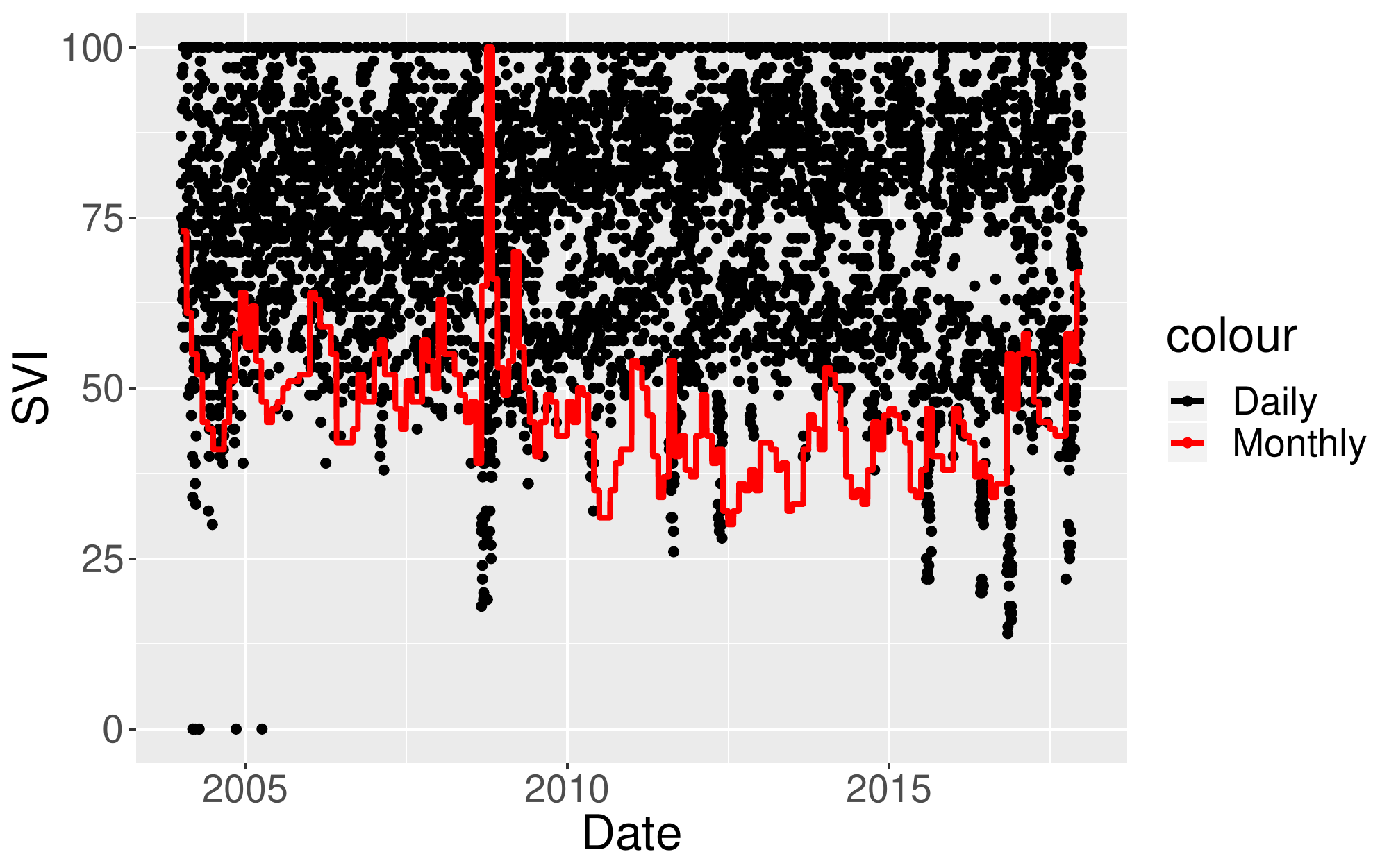}} 
		\subfigure[Adjusted Daily SVI vs. Monthly SVI]{\label{Adjusted}\includegraphics*[width=.44\textwidth]{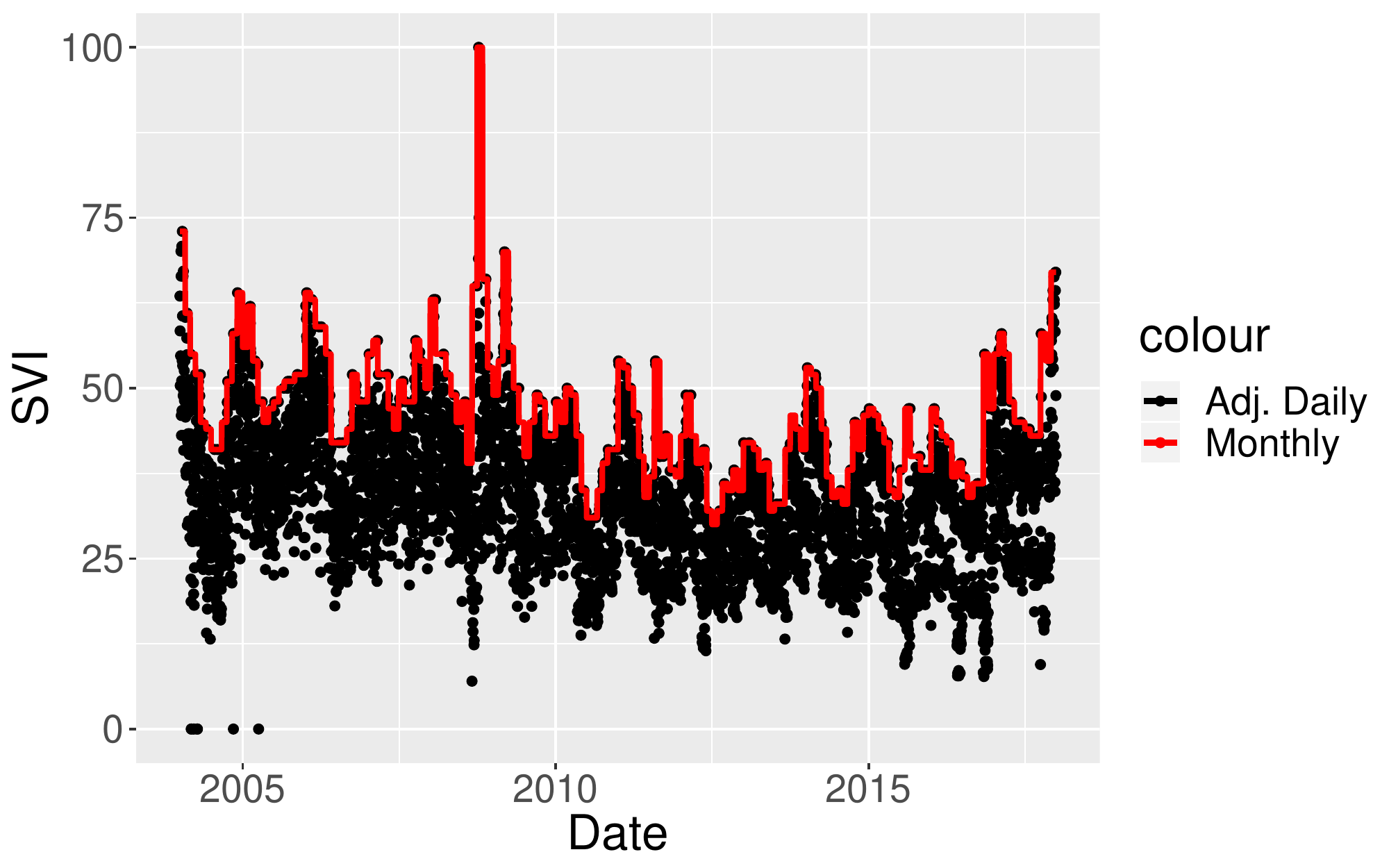}}
		\captionsetup{justification=justified} 
		\caption[Visualization of Daily SVI Adjustment for the term ``bonds"]{Visualization of Daily SVI Adjustment for the term ``bonds" \newline Sub-figure a) depicts the true monthly SVI, as well as the corresponding unadjusted concatenated daily SVI. Sub-figure b) shows the same data after correcting the daily SVI for overall monthly level.} 
	\label{fig:SVI_adjustment_bonds}
	\end{figure}
	
	\begin{figure}[htbp] 
	\centering
	\subfigure[Model (CR)]{\label{Mod1}\includegraphics*[width=.45\textwidth]{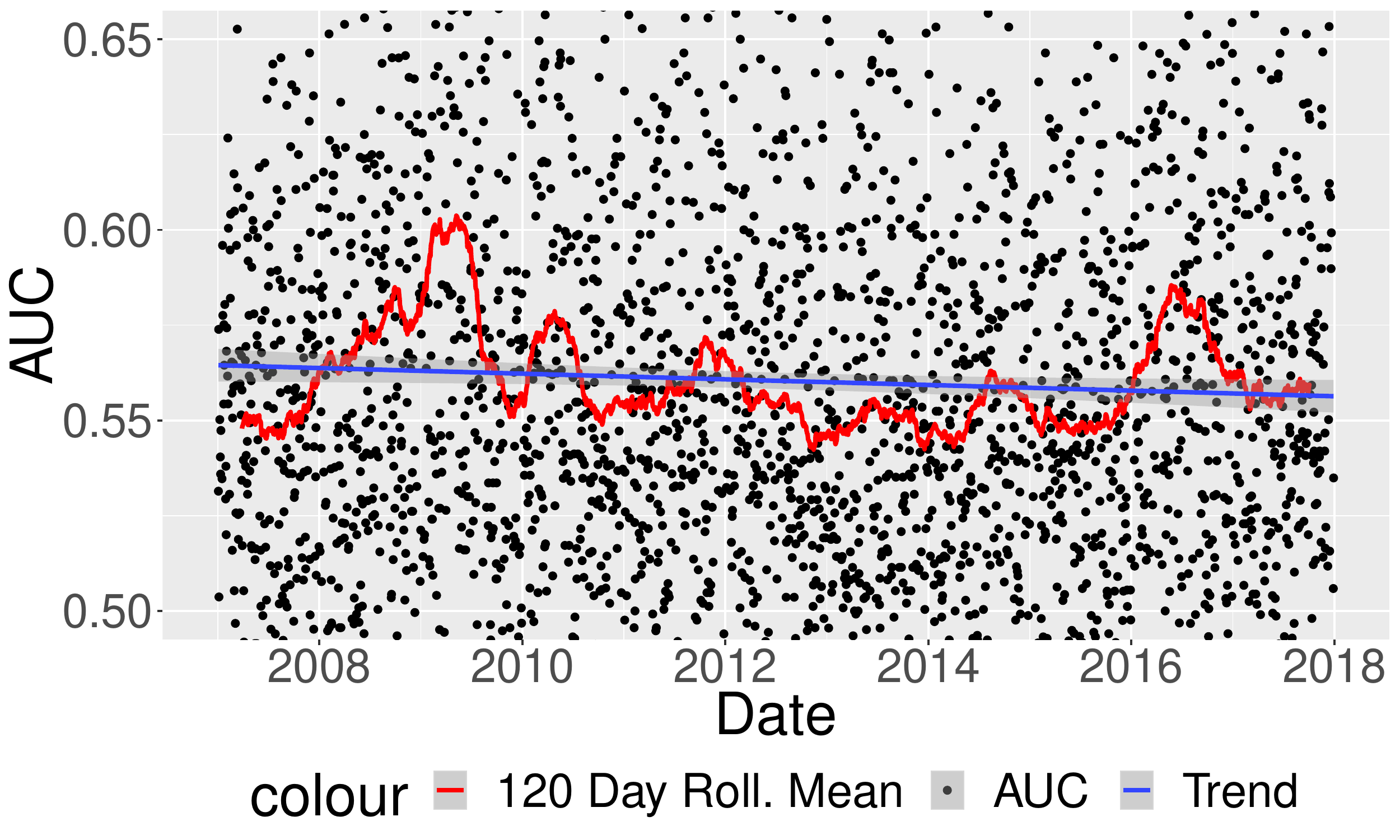}}
	\subfigure[Model (SVI)]{\label{Mod2}\includegraphics*[width=.45\textwidth]{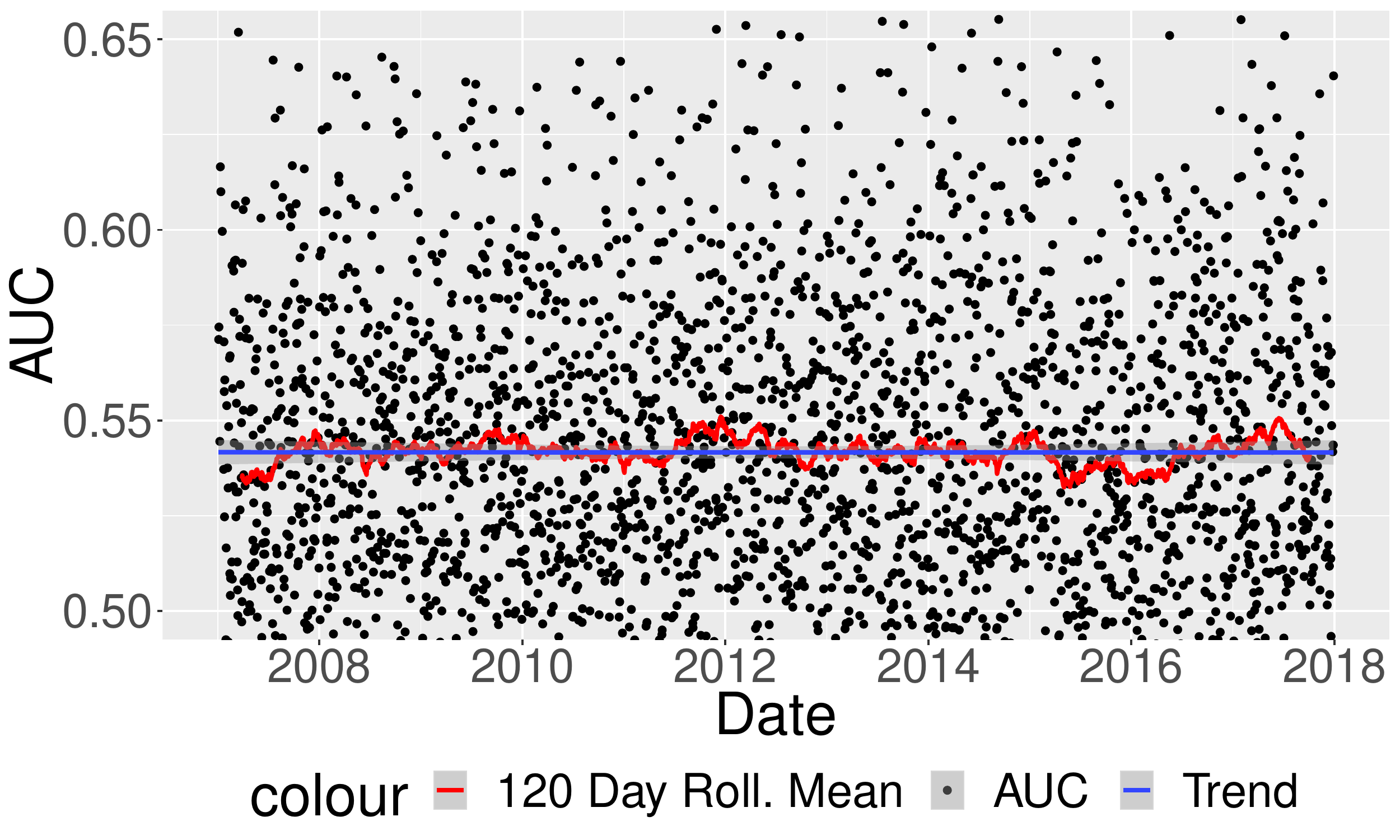}}
	\subfigure[Model (GI)]{\label{Mod3}\includegraphics*[width=.45\textwidth]{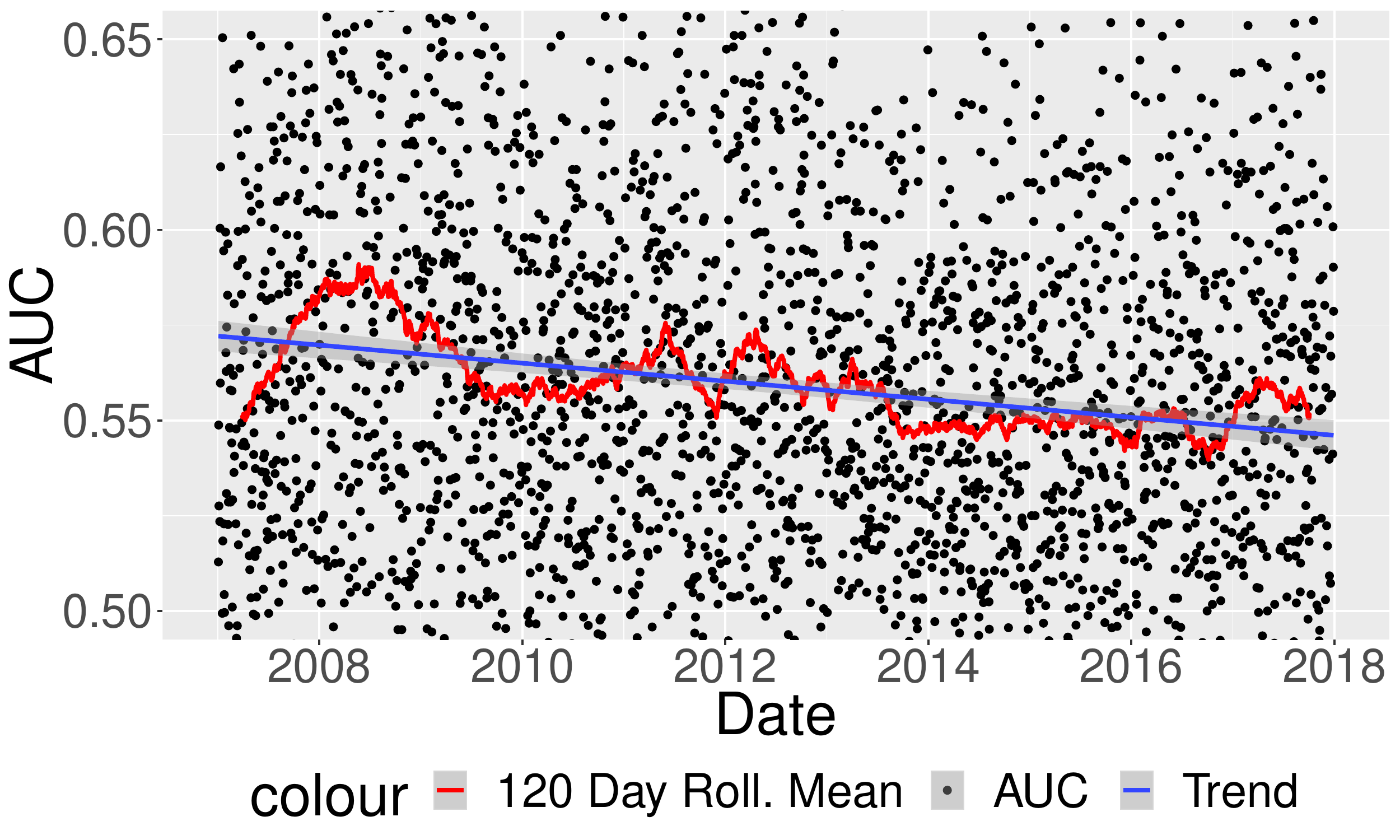}}
	\subfigure[Model (CR, SVI)]{\label{Mod4}\includegraphics*[width=.45\textwidth]{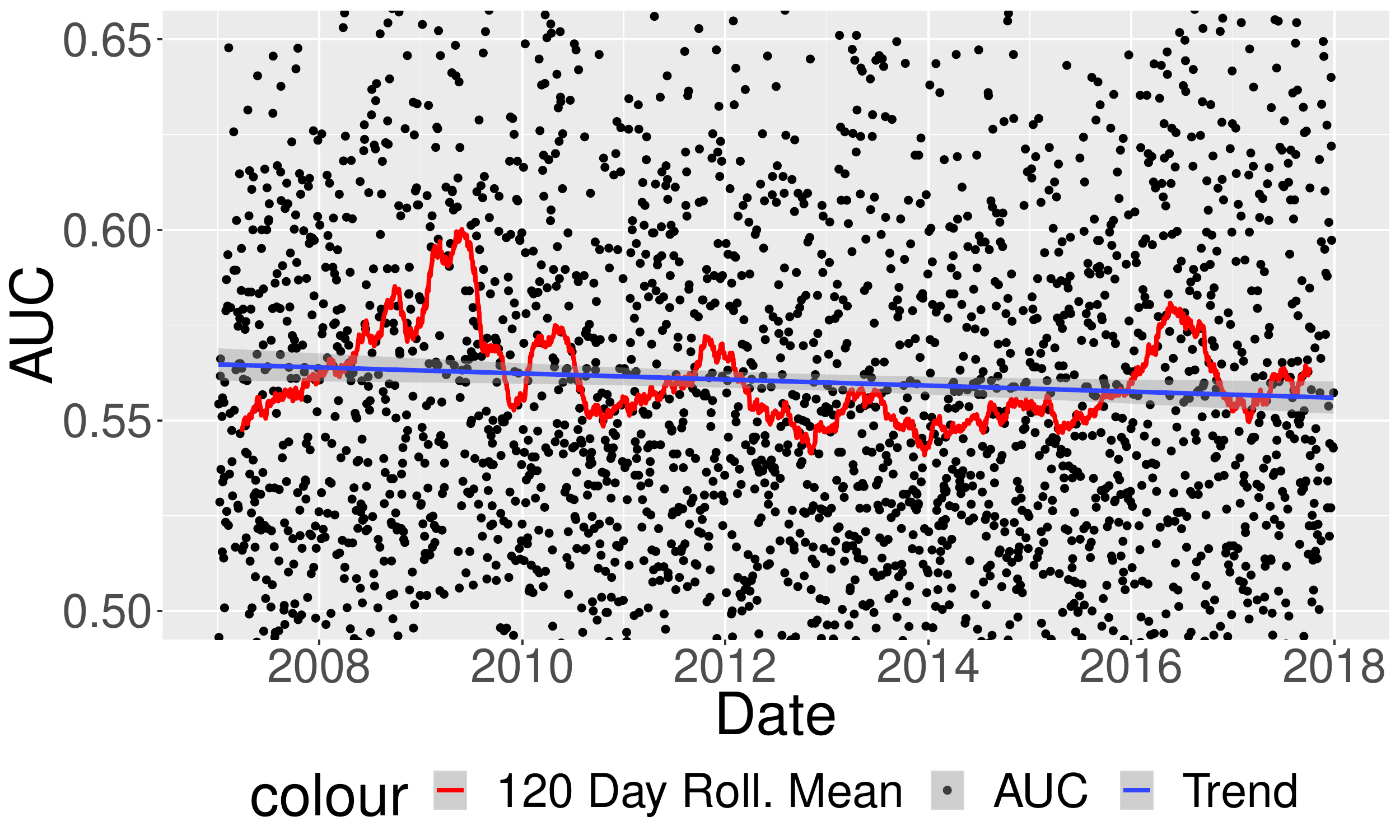}}
	\subfigure[Model (CR, GI)]{\label{Mod5}\includegraphics*[width=.45\textwidth]{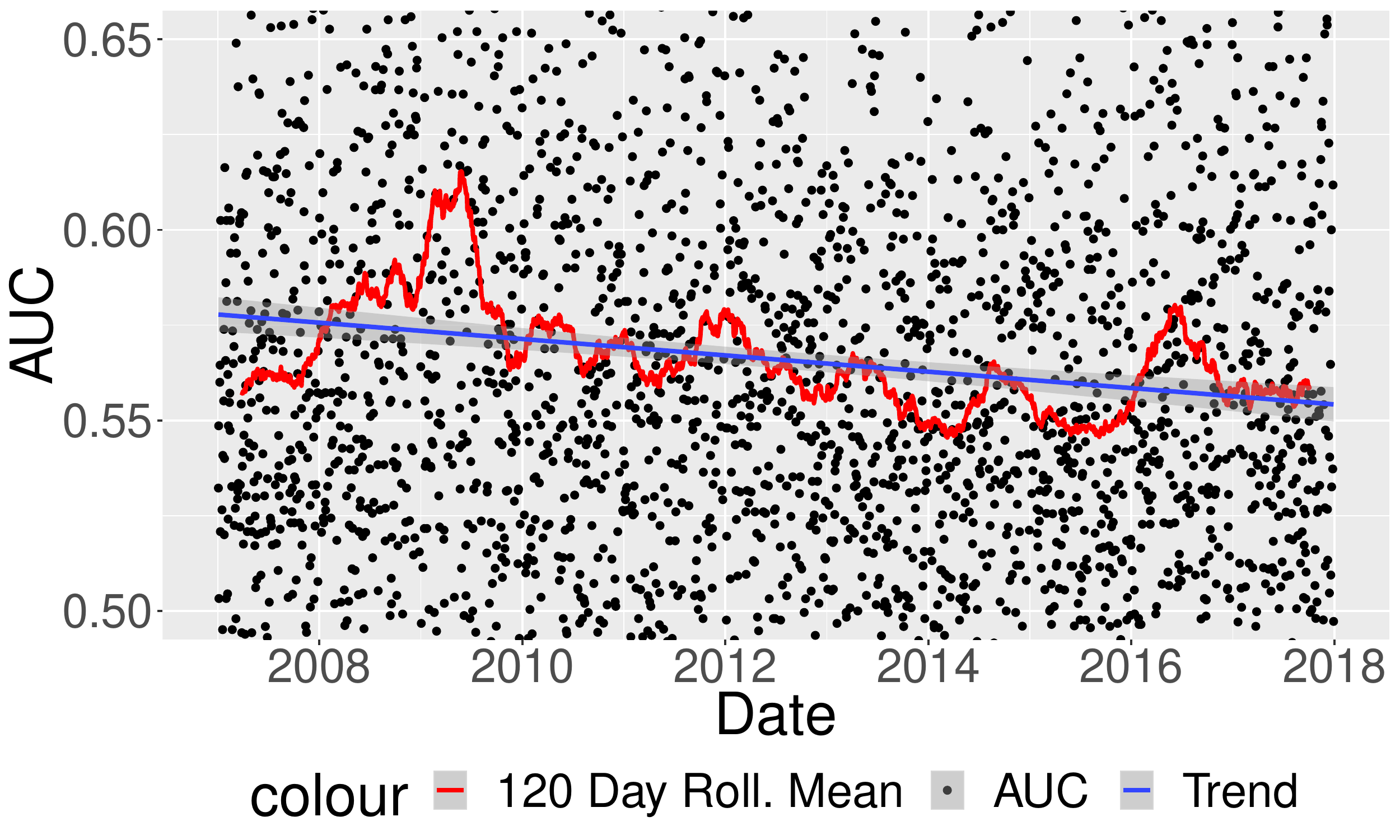}}
	\subfigure[Model (SVI, GI)]{\label{Mod6}\includegraphics*[width=.45\textwidth]{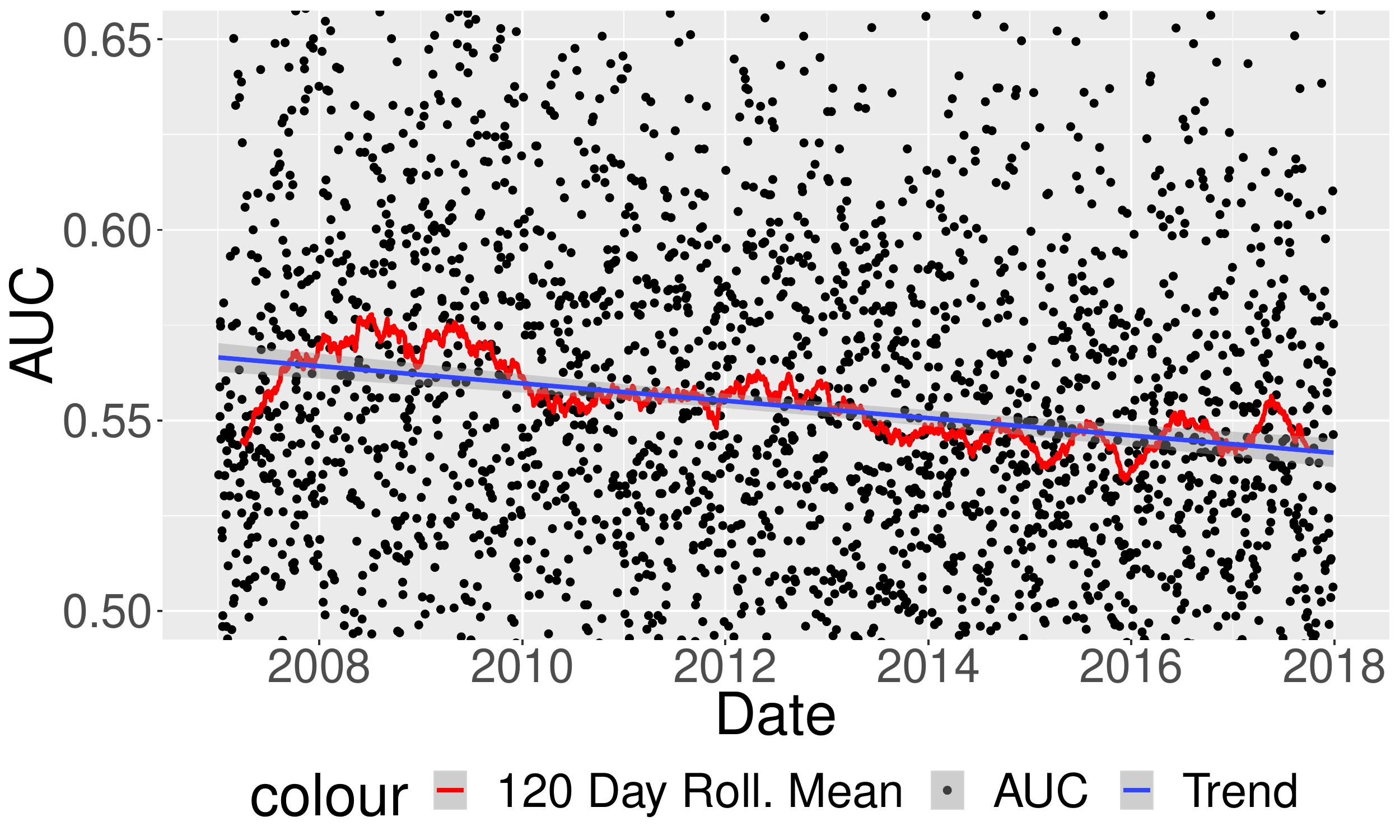}}
	\subfigure[Model (CR, SVI, GI)]{\label{Mod7}\includegraphics*[width=.45\textwidth]{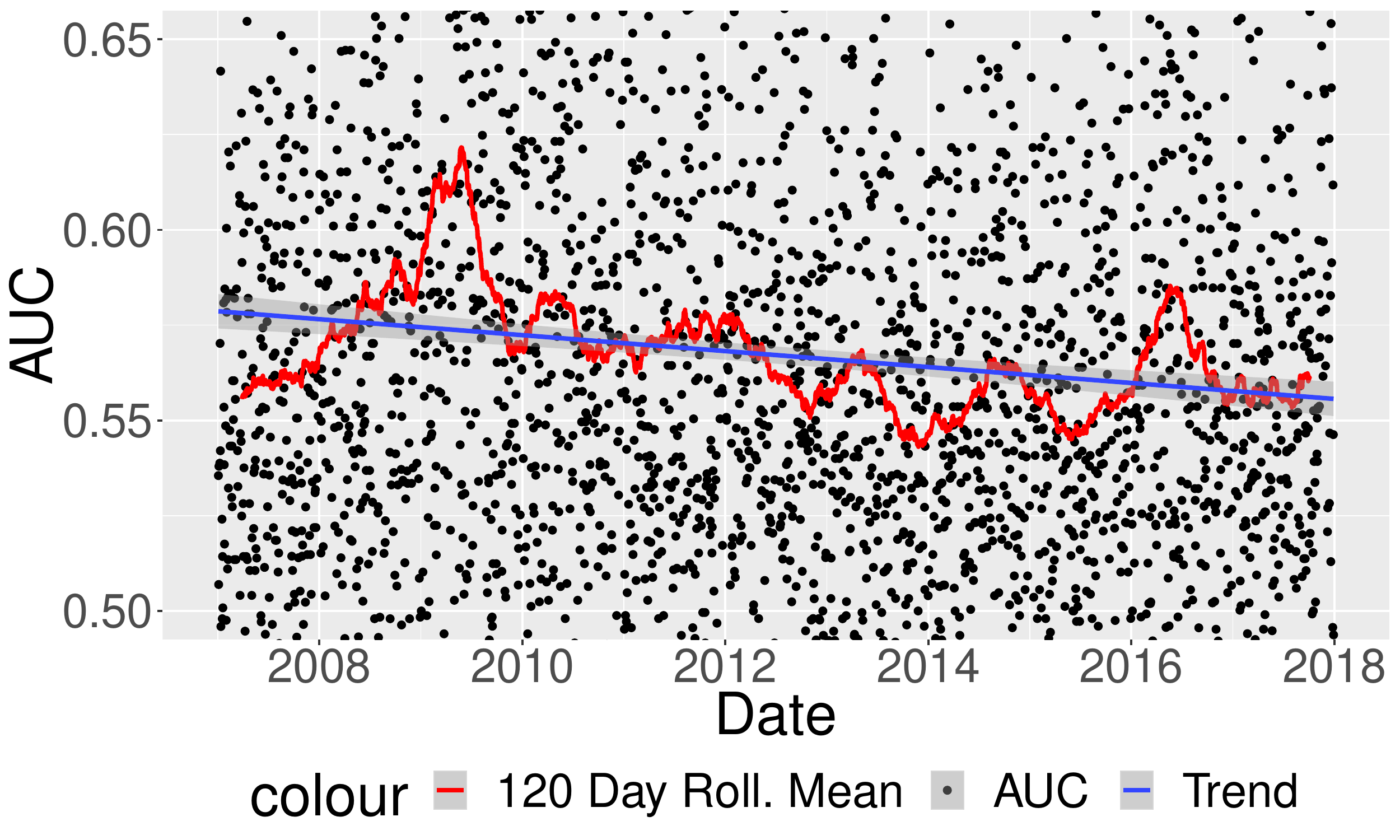}}
	\captionsetup{justification=justified} 
	\caption[Daily AUCs incl. Rolling 120 Days Average and Linear Trend]{Daily AUCs incl. Rolling 120 Days Average and Linear Trend} 
	\label{fig:AverageAUCs}
	\end{figure}
	
	\begin{figure}[htbp] 
	\centering
	\footnotesize
	\includegraphics[width=0.6\textwidth]{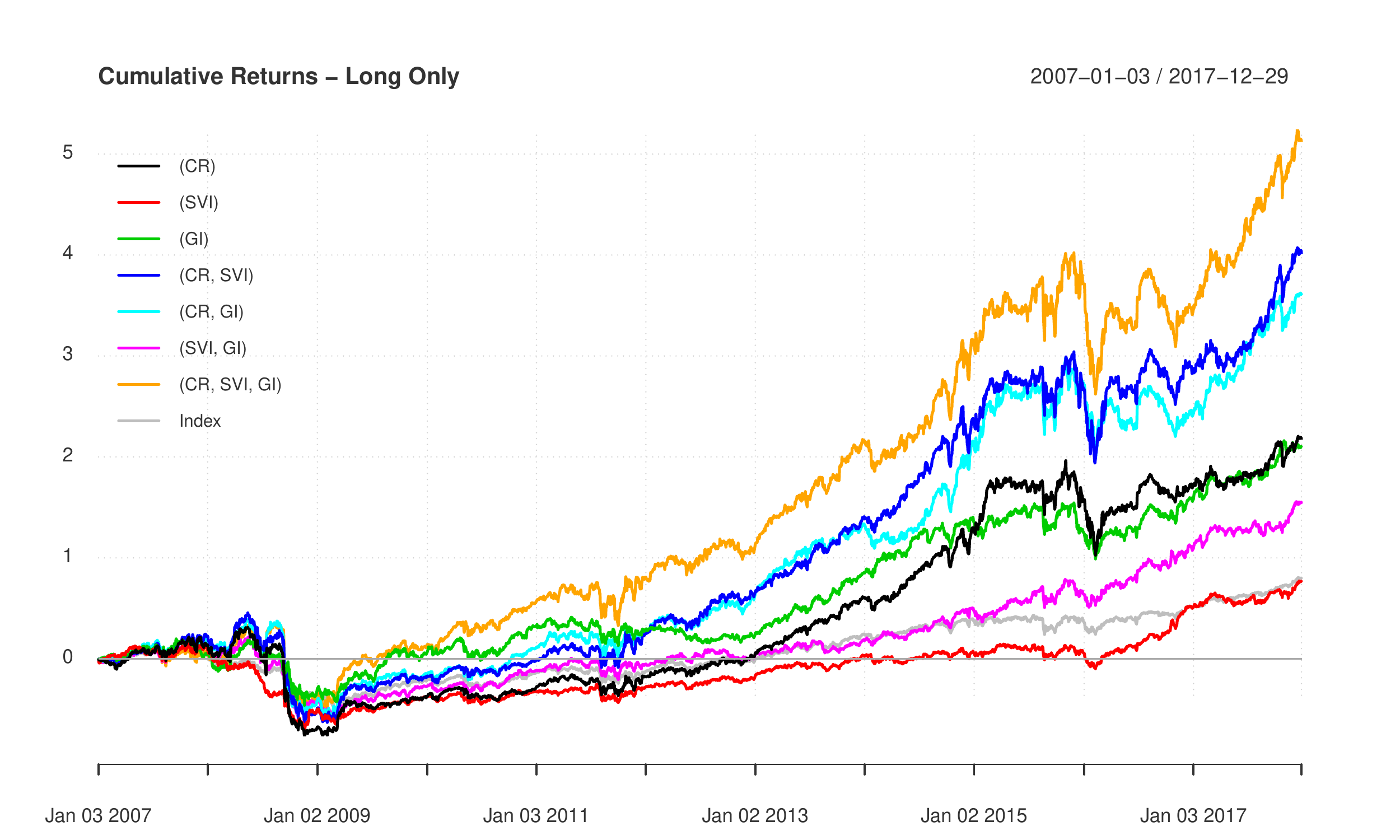}
	\captionsetup{justification=justified} 
	\caption[Cumulative Returns per Model - Long Only]{Cumulative Returns per Model - Long Only \newline The figure depicts the cumulative performance of the stock portfolio consisting daily of the five stocks with the highest probability of directional outperformance of the median return.} 
	\label{fig:Cum_Returns_long}
	\end{figure}
	
	\begin{figure}[htbp] 
	\centering
	\footnotesize
	\includegraphics[width=0.6\textwidth]{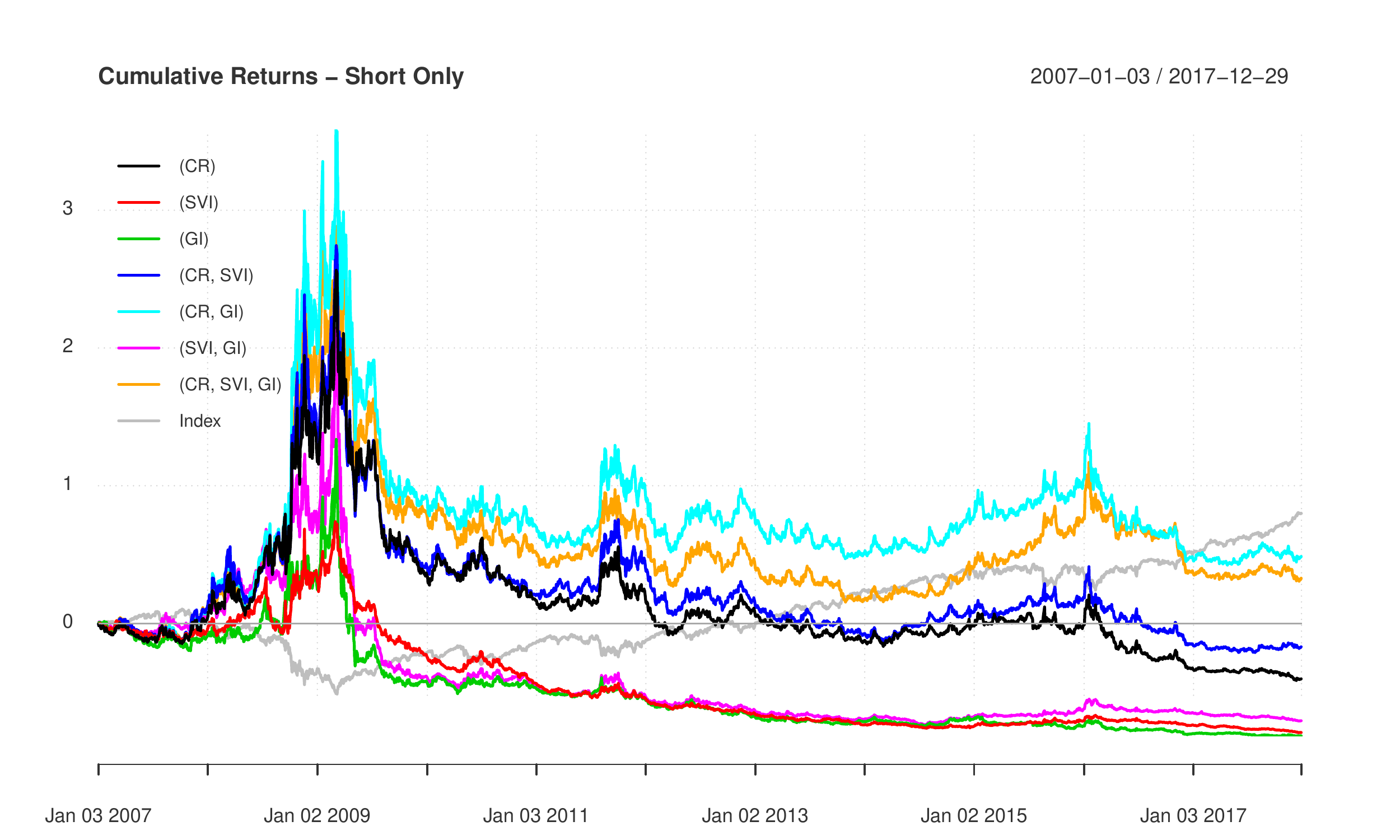}
	\captionsetup{justification=justified} 
	\caption[Cumulative Returns per Model - Short Only]{Cumulative Returns per Model - Short Only \newline The figure depicts the cumulative performance of the stock portfolio consisting daily of the five stocks with the lowest probability of directional outperformance of the median return.} 
	\label{fig:Cum_Returns_short}
	\end{figure}
	
	\begin{figure}[htbp] 
	\centering
	\subfigure[Model (SVI)]{\label{Var2}\includegraphics*[width=.24\textwidth]{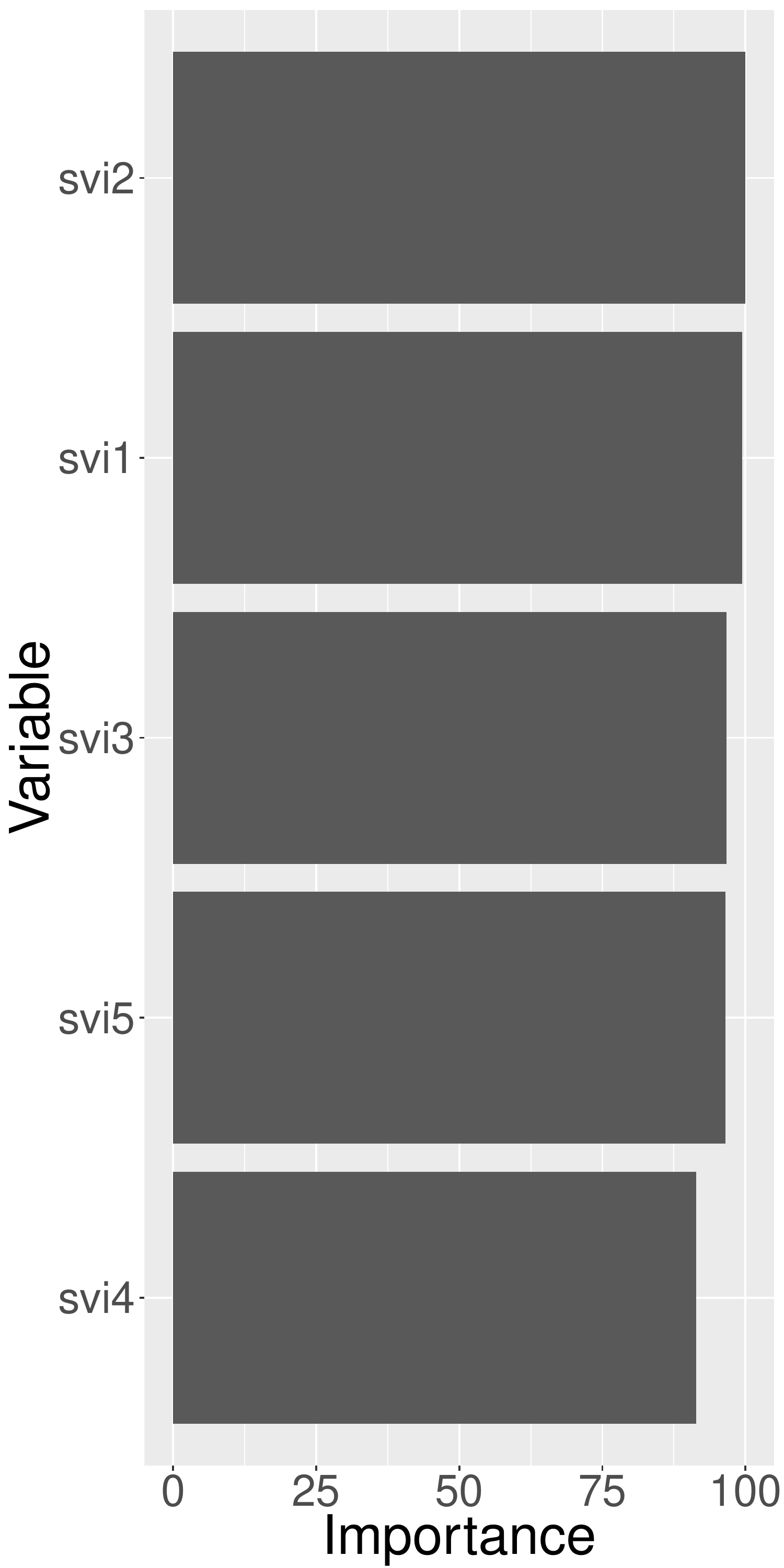}}
	\subfigure[Model (GI)]{\label{Var3}\includegraphics*[width=.24\textwidth]{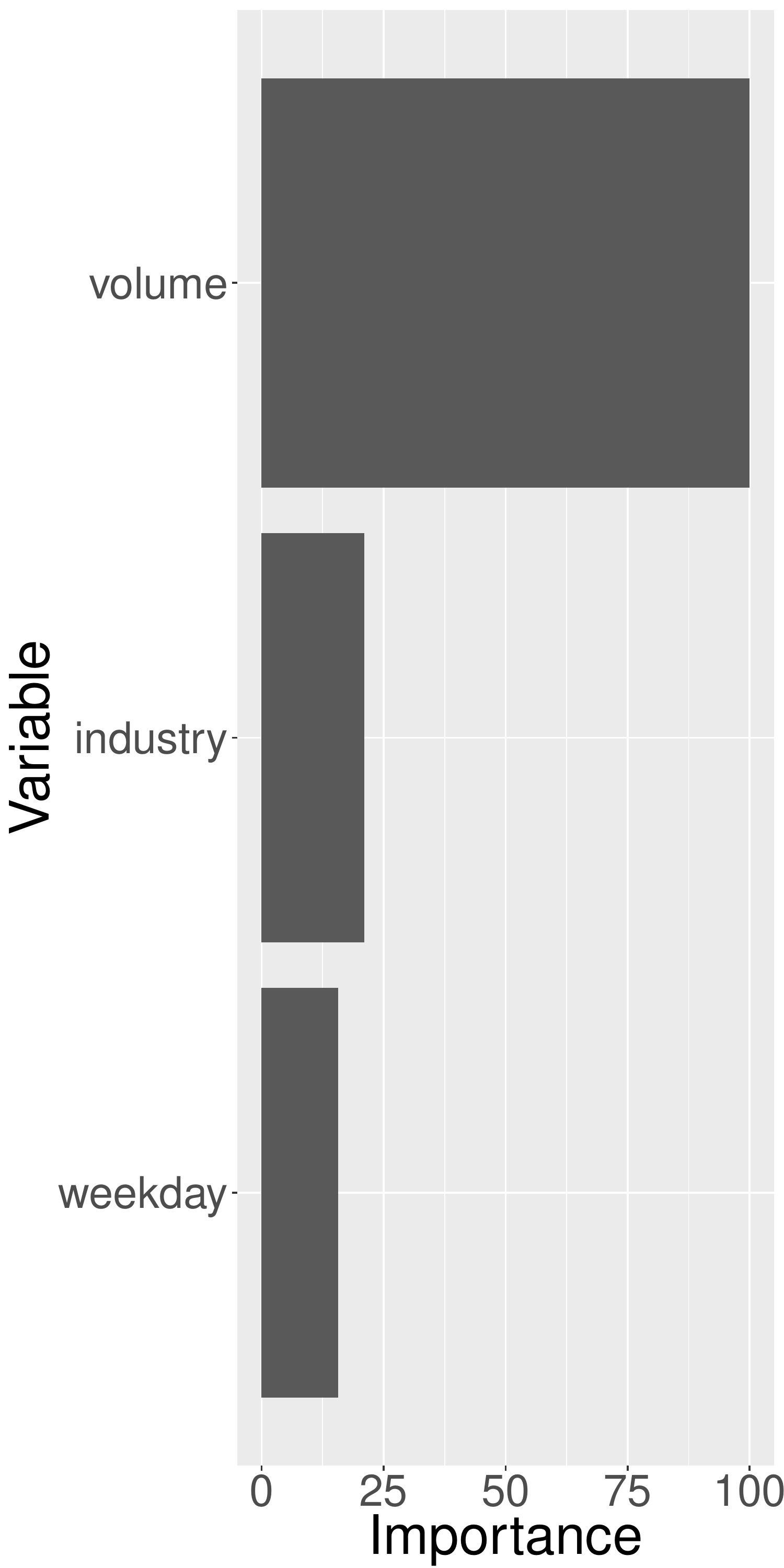}}
	\subfigure[Model (SVI, GI)]{\label{Var6}\includegraphics*[width=.24\textwidth]{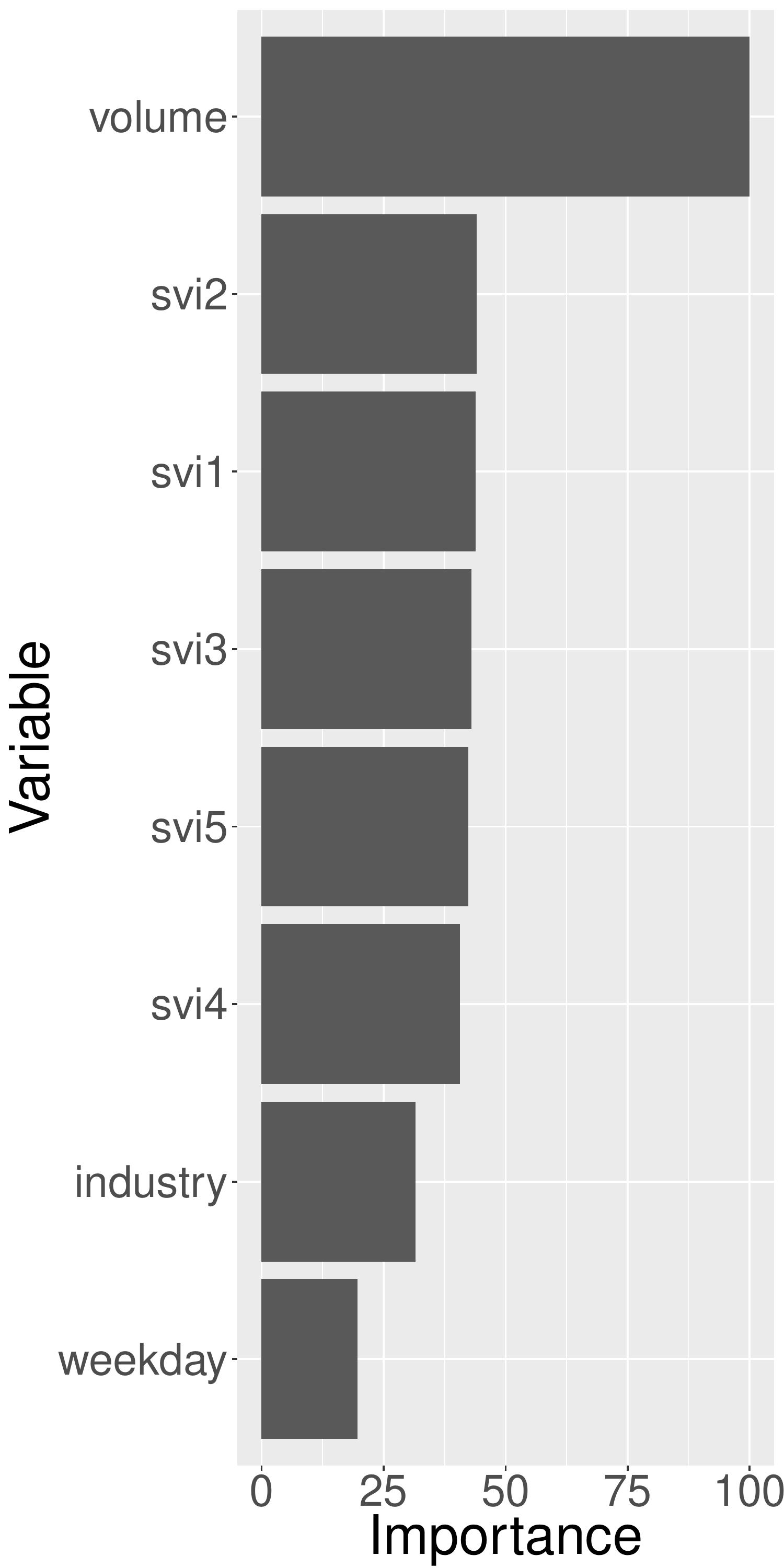}}
	\captionsetup{justification=justified} 
	\caption[Variable Importance per Model]{Variable Importance per Model} 
	\label{fig:VarImportance_236}
	\end{figure}
	
	\begin{figure}[htbp] 
	\centering
	\footnotesize
	\includegraphics[width=0.6\textwidth]{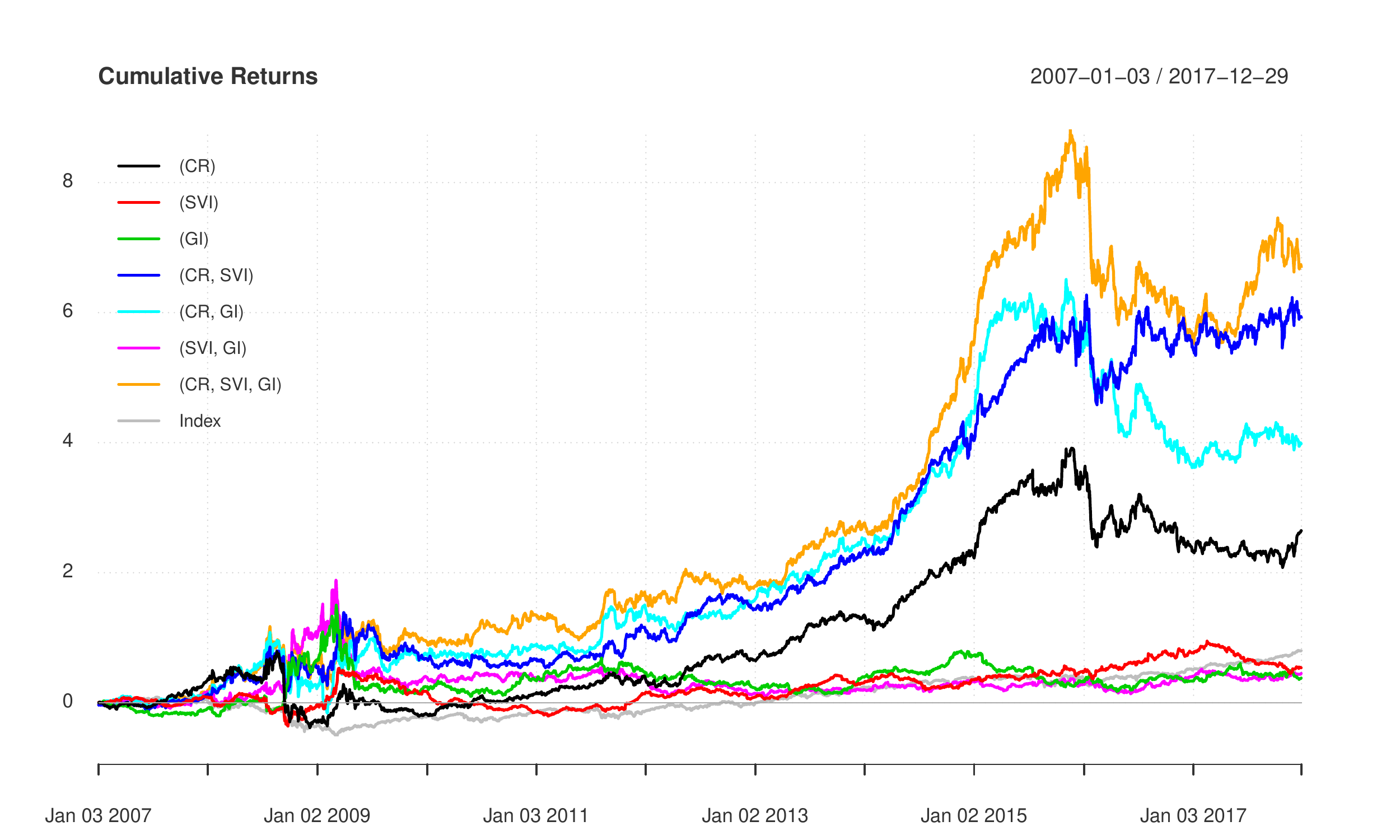}
	\captionsetup{justification=justified} 
	\caption[Cumulative Returns per Data Set with Varying Parametrization]{Cumulative Returns per Data Set with Varying Parametrization \newline For these models the number of boosting iterations is set to 100 and the learning rate is set to 0.1.}
	\label{fig:Cum_Returns_Robustness}
	\end{figure}
\end{appendix}
	
\end{document}